\documentclass[aps,prb,twocolumn, showpacs, superscriptaddress]{revtex4-1}
\usepackage{amsmath, amsthm, amssymb,amsbsy,graphicx,times,subfigure,marvosym,color,bm,hyperref,psfrag,graphics}
\usepackage[latin1]{inputenc}
\usepackage{sidecap}
\usepackage{ulem}
\usepackage{appendix}

\usepackage{comment}
      
\begin{document}

\title{Peering into the darkness: vison-generated photon mass in quantum spin ice}
\author{M. P. Kwasigroch}
\affiliation{London Centre for Nanotechnology, University College London, 17-19 Gordon Street, London WC1H 0AH, U.K.}
\affiliation{T.C.M. Group, Cavendish Laboratory, University of Cambridge, J. J. Thomson Avenue, Cambridge CB3 0HE, U.K.}


\begin{abstract}
  Describing experimental signatures of quantum spin ice has been the focus of many theoretical efforts, as definitive experimental verification of this candidate quantum spin liquid is yet to be achieved. Gapped excitations known as visons have largely eluded those efforts. We provide a theoretical framework, which captures their dynamics and predicts new signatures in the magnetic response. We achieve this by studying the ring-exchange Hamiltonian of quantum spin ice in the large-$s$ approximation, taking into account the compact nature of the emergent $U(1)$ gauge theory. We find the stationary solutions of the action -- the instantons -- which correspond to visons tunneling between lattice sites. By integrating out the instantons, we calculate the effective vison Hamiltonian, including their mass. We show that in the ground state virtual vison pairs simply renormalise the speed of light. At low temperatures, however, thermally activated visons form a Debye plasma and introduce a mass gap in the photon spectrum, equal to the plasma frequency, which we calculate as a function of temperature. We demonstrate that this dynamical mass gap should be visible in energy-resolved neutron scattering spectra but not in the energy-integrated ones. We also show that it does not lead to confinement of static spinons.

\end{abstract}



\maketitle


\section{Introduction}

Quantum spin ice (QSI) is a candidate quantum spin liquid\cite{Gingras}, where an emergent U(1) gauge symmetry prevents magnetic order down to zero temperature and gives rise to gauge and fractionalised excitations: a gapless photon, magnetic monopoles (spinons) and emergent electric charges (visons). In spite of this long list of predicted exotic excitations, definitive experimental verification is still missing, and this situation is shared by QSI with many other proposed spin liquids\cite{Knolle}. Hence, it is important for theoretical efforts to characterise as many signatures of these excitations as possible. In particular, the theoretical description of the inelastic response, through the dynamical structure factor, has recently been successful in providing evidence of a potential spin liquid in $\alpha$-RuCl$_3$\cite{Attila_4,Attila_5, Attila_6, Attila_7, Attila_8, Attila_9}.

In our previous work\cite{Our_Work}, we analytically characterised the photon excitation of QSI by including quantum fluctuations around the classical limit via a large-s expansion. We were also able to calculate the energy of the gapped vison excitation, but were not able to study its dynamics. In this paper, we extend the semiclassical description of our previous work to capture the vison dynamics and its contribution to the inelastic magnetic response of QSI.

The vison is an emergent excitation of QSI, a consequence of the compact U(1) gauge symmetry.\cite{Hermele} (The gauge group is compact because of the quantisation of the spins.) Visons are sources of flux of the emergent electric field. Even though the field is divergenceless, a Dirac string carrying flux that is a multiple of $2\pi$ has zero energy in a compact theory, allowing the charges at its ends to behave as free excitations. The state of these electric charges has profound consequences on the force between the magnetic charges (spinons). In particular, potential condensation of electric charges can lead to the confinement of magnetic charges.

It is believed that at zero temperatures visons simply lead to the renormalisation of the speed of light, but at non-zero temperatures, it is anticipated that their effects are less benign. Ref.~\onlinecite{Gingras} already drew comparisons with (2+1)-dimensional compact lattice gauge theory, which is always in the confined phase and has a gapped photon spectrum\cite{Polyakov_1974}. At increasing temperatures, the (3+1)-dimensional QSI becomes more like a (2+1)-dimensional compact lattice gauge theory and perhaps a condensation of visons leads to a photon mass gap and confinement of magnetic monopoles. 

Despite their importance, so far, few experimental signatures of visons have been predicted\cite{Attila_36, Attila_37}. They are sources of a fictitious electric field and hence cannot easily be probed directly in experiments.  They have also largely eluded theoretical efforts of quantum Monte-Carlo\cite{Shannon, Attila_31, Attila_32, Huang, Attila_34, Attila_35}, due to the limited resolution of excitation spectra, as well as mean-field treatments\cite{Attila_40, Attila_41, Attila_42, Attila_43, Attila_44}. In this context, the work of Ref.~\onlinecite{Attila}, made recent progress by looking at visons through classical Monte Carlo, characterising their effect on emergent field correlators and the heat capacity. This work analysed vison signatures in the classical $s=\infty$ limit. Here, we will show that including non-perturbative corrections in $s$, makes the vison inertia finite and allows us to study their quantum dynamics at low temperatures. We demonstrate that these dynamics show up in the magnetic response, which can be probed directly.

In our theoretical analysis, we  are  motivated  by  the  success  of  our  previous  work\cite{Our_Work},
where we used a large-s semiclassical description to successfully capture the emergent electrodynamics of the ring-exchange Hamiltonian\cite{Hermele} and obtained a photon spectrum that was in good quantitative agreement with quantum Monte Carlo calculations for
$s= 1/2$\cite{Shannon}. As in our previous work, we map the ring-exchange
Hamiltonian in  the  large-$s$
limit  onto  compact  U(1)  lattice
gauge  theory  using  the  Villain  spin  representation.    However, this time we do not neglect spin quantisation, or equivalently, the periodic nature of the emergent dual field, which allows us to include the dynamics of visons in the theoretical description. We show that, in the ground state the effects
of  these  charges  are  rather  innocuous;  they  exist  as  virtual,
tightly bound pairs and simply renormalise the speed of light.
At non-zero temperatures, however, they enter into existence
as physical, thermally activated excitations, that form a plasma.  We show that propagation of the photon through the vison plasma will appear gapped, which can be probed in neutron
scattering experiments.  We calculate typical scattering intensities that might be observed. We note that a similar effect had been anticipated in the context of hot QED by Ref.~\onlinecite{Polyakov_1978}.

The paper is structured as follows. In Sec.~\ref{The_Model} we present the ring-exchange Hamiltonian in the Villain spin representation, which was introduced in our previous work. We then move to the dual electric charge representation, where the visons are made more manifest. We conclude the section by writing down the imaginary-time action of our model and finding its normal modes. In Sec~\ref{Instantons}, we analyse the stationary solutions of the action -- the instantons -- which correspond to the quantum tunneling of visons between neighbouring lattice sites. In Sec.~\ref{Ground_State}, we look at the ground state properties of our model. Sec.~\ref{Non-zero_Temperatures} studies the effects of visons at non-zero temperatures and presents the main results of this paper.  Finally, in Sec.~\ref{Conclusion}, we summarise all our main findings.

\section{The Model} \label{The_Model}

Following on from our previous work\cite{Our_Work}, we study the ring-exchange Hamiltonian of QSI\cite{Hermele} in the Villain spin representation\cite{Villain}

\begin{eqnarray}
\hat{H}&=&g\sum_{\alpha\beta}\Big[ e^{i{\rm curl}_{\alpha\beta} \hat{\phi}/ 2}
\prod_{ij\in\alpha\beta}\left(\tilde{s}^2-\hat{S}_{ij}^z {}^2\right)^{\frac{1}{2}}e^{i{\rm curl}_{\alpha\beta} \hat{\phi}/2}
\nonumber\\
&&+{\rm h.c.} \Big],
\end{eqnarray}
where the spin azimuthal angle $\phi\in(-\pi,\pi]$, its projection on the $z$-axis $S^z$ is an integer or half-integer with $|S^z|\leq s$, $\left[\hat{\phi},\hat{S^z}\right]=i$, $\tilde{s}=s+\frac{1}{2}$, and the product is over all six pyrochlore lattice sites $ij$ belonging to the plaquette centred on site $\alpha\beta$ of the {\it dual} pyrochlore lattice. The curl is taken around this plaquette. Just in like our previous work, Latin letters $\{i\}$ index the sites of the diamond lattice (bond midpoints $\{ij\}$ correspond to pyrochlore lattice sites on which the spins live) and the Greek letters index the sites of the dual diamond lattice (bond midpoints $\{\alpha\beta\}$ correspond to the dual pyrochlore lattice sites on which the plaquettes are centred). The spins satisfy the constraint
\begin{eqnarray}
{\rm div}_i S^z \equiv\sum_{ij}\nolimits ^{(i)}S^z_{ij}=Q_i,
\label{eq: S^z constraint}
\end{eqnarray}
where the sum is taken over the four pyrochlore lattice sites $ij$ that belong to the diamond lattice site $i$, i.e. over the four corners of the tetrahedron centred on $i$ (note that a positive sign is taken for 'up' tetrahedra and negative for 'down' tetrahedra). $Q_i \in \mathbb{Z}$ are static magnetic charges (magnetic monopoles) introduced into the system. The charges are static because the constraint commutes with the Hamiltonian and $Q_i$ are therefore constants of motion.

\subsection{Electric charge representation}
The visons are made more manifest in the dual, electric charge representation of the Hamiltonian.
Rather than working with the conjugate magnetic field $\hat{S}^z_{ij}$ and the electric vector potential $\hat{\phi}_{ij}$, we shall be working with the conjugate electric field and the magnetic vector potential. We introduce the new conjugate operators $\{\hat{E}_{\alpha\beta},\hat{A}_{\alpha\beta}\}$ via
\begin{eqnarray}
\hat{E}_{\alpha\beta} &=& 
{\rm curl}_{\alpha\beta}\hat{\phi},\label{magnetic field}
\\
\hat{S}^z_{ij}&=&{\rm curl}_{ij}\hat{A}+\vartriangle_{ij}\psi,\label{Helmholtz}
\end{eqnarray}
where $\psi_i$ is a scalar field (not an operator) defined on diamond lattice sites with $\vartriangle_{ij} \psi  \equiv\psi_j-\psi_i$. Because the electric field $E_{\alpha\beta}$ is defined as the lattice curl it has zero divergence.  Eq.~\eqref{Helmholtz} is just the lattice Helmholtz decomposition for the magnetic field and because its divergenceful part is a constant of motion, it can be expressed as $\vartriangle_{ij}\psi$ with $\psi_i$ a scalar field.  The commutation relations for the magnetic $\hat{S^z_{ij}}$  and electric field
$\hat{E}_{\alpha \beta}$ operators (which follow from $[\hat{\phi}_{ij},S^z_{ij}]=i$) imply that the new conjugate operators have the canonical commutator $[\hat{E}_{\alpha \beta}, \hat{A}_{\alpha \beta}]=-i$. 

 Because the Hamiltonian is periodic in $\hat{E}_{\alpha\beta}$, the non-integer part of $\hat{A}_{\alpha\beta}$, $A^0_{\alpha\beta}$ is a constant of motion and is analogous to crystal momentum. It is therefore useful to make the following replacement
\begin{eqnarray}
\hat{A}_{\alpha\beta} \rightarrow \hat{A}_{\alpha\beta} + A^0_{\alpha\beta},
\end{eqnarray}
where the new operators $\hat{A}_{\alpha\beta}$ have strictly integer eigenvalues. The energy eigenstates are now given by
\begin{eqnarray}
|\Psi\rangle
&=&
\sum_{A_{\alpha\beta}} 
c_{A_{\alpha\beta}}
|A_{\alpha\beta}\rangle \otimes |A^0_{\alpha\beta} \rangle
\nonumber\\
&=&
\int d{E_{\alpha\beta}} \;
\Psi(E_{\alpha\beta})
|E_{\alpha\beta}\rangle \otimes |A^0_{\alpha\beta} \rangle,
\end{eqnarray}
where $\Psi(E_{\alpha\beta})$ is a periodic function of $E_{\alpha\beta}$ and the eigenstates are analogous to Bloch states. (The kets are eigenstates of the respective operators.)
 After the above replacement, the magnetic field becomes
\begin{eqnarray}
\hat{S}^z_{ij}&=&{\rm curl}_{ij}\hat{A}_{\alpha\beta}+B^{0}_{ij},\nonumber\\
B^{0}_{ij} &=& {\rm curl}_{ij}A^0_{\alpha\beta}+\vartriangle_{ij}\psi,
\label{eq: background magnetic field}
\end{eqnarray}
where $B^{0}_{ij}$ is a {\it static} background field. Different constraints on the allowed values of the magnetic field $\hat{S}^z_{ij}$ can be implemented as constraints on the allowed eigenstates of $\hat{A}_{\alpha\beta}$ and values of the static background field $B^0_{ij}$. For instance, to realise the constraint that the magnetic field is half-integer valued we simply choose any background field $A^0_{\alpha\beta}$ that is  half-integer valued. The field $\hat{A}_{\alpha\beta}$ then becomes unconstrained.  The kinematic constraint $|S^z_{ij}|\leq s$ is the hardest to implement, since for any background field  $A^0_{\alpha\beta}$, there are forbidden eigenstates of $\hat{A}_{\alpha\beta}$. As discussed in our previous work~\cite{Our_Work}, in the large-$s$ limit, the typical fluctuations $S^z_{ij}\sim \sqrt{s}$ and hence this constraint is largely irrelevant.

Any magnetic charges $Q_i$ introduced into the system (see Eq.~\ref{eq: S^z constraint}) uniquely determine $\psi_i$ via the lattice Laplace equation
\begin{eqnarray}
{\rm div}_i B^{0} \equiv \nabla^2_i \psi = Q_i, 
\end{eqnarray}
where $\nabla^2_i$ is the lattice Laplacian. To realise the constraint that the magnetic field is integer (half-integer) valued, we must then choose a configuration $A^0_{\alpha\beta}$ that makes $B^{0}_{ij}$ integer ( half-integer) -- this is important for confinement and is analysed in detail in appendix~\ref{App_Confinement}. 

Because energy eigenstates are periodic in $E_{\alpha\beta}$, without loss of generality, we will restrict $E_{\alpha\beta} \in (-\pi,\pi]$, i.e. measure the field modulo $2\pi$. The electric field now acquires a non-zero divergence quantised in units of $2\pi$
\begin{eqnarray}
{\rm div}_{\alpha}\hat{E} \equiv \sum\nolimits_{\alpha\beta}^{(\alpha)} \hat{E}_{\alpha \beta} =2\pi q_{\alpha},
\label{eq: B quantisation}
\end{eqnarray}
where $q_{\alpha}\in \mathbb{Z}$ are the emergent electric charges ($(\alpha)$ identifies that the sum is taken over the four corners of the tetrahedron $\alpha$ of the dual diamond lattice and again positive sign is taken for 'up' tetrahedra and negative for 'down' tetrahedra). A single electric charge at the tetrahedron $\alpha$ ($q_{\alpha}=\pm 1$) gives rise to a smooth, long-wavelength modulation of the electric field $E_{\alpha\beta} \propto \frac{1}{R^2}$, where $R$ is the distance from the electric charge. This long-wavelength modulation is a gapped topological excitation known as the vison.

To summarise this subsection, we write down the ring-exchange Hamiltonian in the electric charge representation 
\begin{eqnarray}
\hat{H}&=&-g\sum_{\alpha\beta}\Big\{
 e^{i\hat{E}_{\alpha\beta}/2}
\prod_{ij\in\alpha\beta}  \left[  \tilde{s}^2- \left({\rm curl}_{ij}\hat{A}+B^0_{ij}\right)^2  \right]^{\frac{1}{2}}
\nonumber\\
&&\times e^{iE_{\alpha\beta}/2}
+{\rm h.c.} \Big\}.
\end{eqnarray}
We will be working in the $s\gg 1$ limit and expand the Hamiltonian accordingly
\begin{eqnarray}
\hat{H}&=&\tilde{g}\sum_{\alpha\beta}
 \hat{E}_{\alpha\beta}^2
 +\sum_{ij}\frac{\tilde{g}z}{s^2}\left({\rm curl}_{ij}\hat{A}+B^0_{ij}\right)^2 
 \nonumber\\
 &&+ \mathcal{O}(s^{-2})
 ,
  \label{eq: expanded Hamiltonian}
\end{eqnarray}
where $\tilde{g} = gs^6$ and $z=6$. We have also replaced $\tilde{s}$ with $s$, as we will be working to the first non-vanishing order in $s$ for all physical quantities that we calculate.

\subsection{Partition function}
To study the non-perturbative effects of dynamical electric charges, we begin with the partition function for the above Hamiltonian, obtained in the usual way by inserting two resolutions of the identity in the $\hat{E}_{\alpha\beta}$ and $\hat{A}_{\alpha\beta}$ bases into each of the $N_{\tau}$ Suzuki-Trotter time slices (the time slice width is given by $\epsilon=\frac{\beta }{N_{\tau}}$).

\begin{eqnarray}
&&\mathcal{Z} =
\prod_{\tau, \alpha\beta,\gamma} \int_{-\pi}^{ \pi} d E_{\alpha\beta}  
(\tau) \int_{-\infty}^{+\infty}d A_{\alpha\beta} (\tau)\int_{-\infty}^{+\infty} d \varphi_{\gamma}(\tau) \nonumber\\
&&\times \sum_{j_{\alpha\beta}(\tau)} e^{-\mathcal{S}}  \label{partition and action}
\, ,
\nonumber\\
&&\mathcal{S}
= \sum_{\tau} \Big\{  i\sum_{\alpha\beta}A_{\alpha\beta}(\tau)\vartriangle_{\tau}E_{\alpha\beta}(\tau)
+ \epsilon\tilde{g}\sum_{\alpha\beta}E_{\alpha\beta}^2(\tau)
\nonumber\\
&&
+\epsilon\frac{\tilde{g}z}{s^2} \sum_{ij}\left[  {\rm curl}_{\ij}A(\tau) + B^0_{ij}\right] ^2 \nonumber\\
&&
+\epsilon\sum_{\gamma}i\varphi_{\gamma}(\tau)\left[{\rm div}_{\gamma}E(\tau)-2\pi q_{\gamma}(\tau) \right] \Big\} \nonumber\\
&&+\sum_{\tau, \alpha\beta}i2\pi j_{\alpha\beta}(\tau)A_{\alpha \beta}(\tau),
\label{Action}
\end{eqnarray}
where $\vartriangle_{\tau}E_{\alpha\beta} \equiv E_{\alpha\beta}(\tau + \epsilon) - E_{\alpha\beta}(\tau )$, the sum over integers $j_{\alpha\beta}(\tau)$ ensures that $A_{\alpha\beta}(\tau)$ are integer-valued, and $\varphi_{\gamma} (\tau)$ are Lagrange multipliers that ensure the divergence of the electric field at each dual diamond lattice site $\gamma$ is an integer multiple $q_{\gamma}$ of $2\pi$ for all $\tau$. The Lagrange multipliers can be interpreted as the scalar electric potential.

Note that the zero-modes, arising from the U(1) gauge symmetry of the action, $A_{\alpha\beta}(\tau)\rightarrow A_{\alpha\beta}(\tau) + \chi_{\beta}(\tau)-\chi_{\alpha}(\tau)$, $\varphi_{\gamma}(\tau)\rightarrow \varphi_{\gamma}(\tau)+\chi_{\gamma}(\tau+\epsilon)-\chi_{\gamma}(\tau)$ enforce vison charge conservation
\begin{eqnarray}
\vartriangle_{\tau}q_{\gamma}+ {\rm div}_{\gamma}j=0,\label{continuity}
\end{eqnarray}
for all tetrahedra $\gamma$. We can see that $j_{\alpha\beta} (\tau)$ can be interpreted as vison currents between the tetrahedra touching at $\alpha\beta$.

\subsection{Normal modes of the action
\label{sec: normal modes}}

We first conveniently parametrise the sites of the dual pyrochlore lattice.
The dual pyrochlore lattice is a  superposition of four fcc lattices indexed by $\mu=1,2,3,4$. Taking a single 'up' tetrahedron from the dual diamond lattice, with its centre located at the position vector $\mathbf{r}$, the position vectors of its four corners ($\mu=1,2,3,4$) are $\left(\mathbf{r}+\mathbf{e_{\mu}}/2\right)$, where the four basis vectors $\mathbf{e}_{\mu}$ are given by
\begin{eqnarray}
&&\mathbf{e_1}= \frac{a_0}{4}(1,1,1), \:\: \mathbf{e}_2=\frac{a_0}{4}(1,-1,-1)\nonumber\\
&&\mathbf{e}_3=\frac{a_0}{4}(-1,1,-1), \:\: \mathbf{e}_4=\frac{a_0}{4}(-1,-1,1).
\end{eqnarray}
The centres of all 'up' tetrahedra map out an fcc lattice. Each $\mu$ fcc lattice is then a translation of this lattice by $\mathbf{e_{\mu}}/2$ and corresponds to the set of all those 'up' tetrahedra corners that are displaced by $\mathbf{e_{\mu}}/2$ from their centres. We thus identify each dual pyrochlore lattice site $\alpha\beta$ by an index $\mu$, corresponding to the fcc lattice to which this site belongs, and its position vector on that fcc lattice.  This is reflected in the following change of notation for the variables
\begin{eqnarray}
E_{\alpha\beta}(\tau)  \rightarrow  E_{\mu}(\mathbf{r}_{\alpha}+\mathbf{e_{\mu}}/2,\tau),
\nonumber\\
A_{\alpha\beta}(\tau)  \rightarrow  A_{\mu}(\mathbf{r}_{\alpha}+\mathbf{e_{\mu}}/2,\tau),
\nonumber\\
j_{\alpha\beta}(\tau) \rightarrow j_{\mu}(\mathbf{r}_{\alpha}+\mathbf{e_{\mu}}/2,\tau),
\end{eqnarray}
where  $\mathbf{r}_{\alpha}+\mathbf{e_{\mu}}/2$ is the position vector of the site $\alpha\beta$ and $\mu$ identifies the fcc lattice to which it belongs. $\mathbf{r}_{\alpha}$ and $\mathbf{r}_{\beta}$ are the position vectors of the 'up' and 'down' tetrahedra, respectively, which touch at the site $\alpha\beta$, and $\mathbf{e}_{\mu}=\mathbf{r}_{\beta} - \mathbf{r}_{\alpha}$. 

 We also take the continuum limit $\epsilon\rightarrow 0$ of $\mathcal{S}$, where
\begin{eqnarray}
&&\sum_{\tau} \epsilon \rightarrow \int_{-\beta/2}^{\beta/2} d \tau,
\nonumber\\
&&\frac{\vartriangle_{\tau} E_{\alpha\beta}}{\epsilon} \rightarrow \dot{E}_{\alpha\beta},
\nonumber\\
 &&\sum_{\tau} j_{\alpha\beta} (\tau) A_{\alpha\beta} (\tau) 
\rightarrow
\nonumber\\
&&\int_{-\beta/2}^{\beta/2} d\tau A_{\alpha\beta} (\tau) 
\left[ \sum_{\tau_0} j_{\alpha\beta}(\tau_0)  \delta(\tau - \tau_0) \right],
\end{eqnarray}
 and we will set the background field $E^0_{\alpha\beta}$ to zero for now and return to analysing its effects later.

Because $A_{\alpha\beta}(\tau)$ is a continuous variable, provided we enforce charge conservation in Eq.~\ref{continuity}, we can fix its gauge via the usual Faddeev-Popov procedure\cite{Faddeev-Popov}. We choose to work in the Coulomb gauge
\begin{eqnarray}
{\rm div}_{\alpha} A \equiv \sum\nolimits_{\alpha\beta}^{(\alpha)} A_{\alpha\beta} =0,
\end{eqnarray}
where the longitudinal and transverse parts of the electric field $E_{\alpha\beta}(\tau)$ decouple. (The longitudinal part of $E_{\alpha\beta}$ satisfies ${\rm curl}_{ij}E (\tau)=0$ everywhere, whereas the transverse part satisfies ${\rm div}_{\alpha} E (\tau)=0$ everywhere.) In the Coulomb gauge, the action can be decomposed as follows
\begin{eqnarray}
\mathcal{S} = \mathcal{S}_{\rm long.} + \mathcal{S}_{\rm tran.},
\end{eqnarray}
where $\mathcal{S}_{\rm long.}$ is a functional of the longitudinal part of $E_{\alpha\beta}(\tau)$ only, and $\mathcal{S}_{\rm tran.}$ of the transverse part only. 

\subsubsection{Transverse modes}

The transverse part of the action $\mathcal{S}_{\rm tran.}$ is diagonal in the eigenbasis of the ${\rm curl}$ operator, which is defined as
\begin{eqnarray}
{\rm curl}_{ij}A
&\equiv&
 \sum_{\nu\neq\mu,\pm}
 \pm A_{\nu}(\mathbf{r}_{i}+\mathbf{e}_{\mu}/2 \pm\mathbf{\Delta}_{\mu \nu}, \tau), 
\label{eq: curl in real space}
\end{eqnarray} 
where \begin{eqnarray}
\mathbf{\Delta}_{\mu \nu} \equiv \frac{a_0}{\sqrt{8}}\frac{\mathbf{e}_{\mu}\times \mathbf{e}_{\nu}}{|\mathbf{e_{\mu}}\times \mathbf{e_{\nu}}|} ,
\end{eqnarray}
and $\mathbf{r}_{i}+\mathbf{e}_{\nu}/2$ is the position vector of the plaquette centre $ij$ on the original pyrochlore lattice. The decomposition for the original pyrochlore lattice works in the same way as that for the dual pyrochlore lattice: $\mathbf{r}_{i}$ gives the position vector of the centre of the $i$ 'up' tetrahedron, of the original diamond lattice, and centres of all 'up' tetrahedra map out an fcc lattice, which can be translated by $\mathbf{e}_{\nu}/2$ to give one of the four $\nu=1,2,3,4$ fcc lattices the plaquette centre $ij$ belongs to.

We transform to the eigenbasis of the ${\rm curl}$ operator in the same way that we have done in our previous work\cite{Our_Work} for the original pyrochlore lattice
\begin{eqnarray}
&&A_{\mu}(\mathbf{r}+\mathbf{e}_{\mu}/2,\tau) =
\nonumber\\
 &&\frac{1}{\sqrt{N_s \beta}}
\sum_{\mathbf{k} \in {\rm BZ},\omega, \lambda}U^{\dagger}_{\mu\lambda}(\mathbf{k})A_{\lambda}(\mathbf{k},\omega)
e^{i\mathbf{k}\cdot \left( \mathbf{r}+\mathbf{e}_{\mu}/2 \right)  -i \omega \tau},
\nonumber\\
\nonumber\\
&&E_{\mu}(\mathbf{r}+\mathbf{e}_{\mu}/2,\tau) =
\nonumber\\
&&\frac{1}{\sqrt{N_s \beta}}
\sum_{\mathbf{k} \in {\rm BZ}, \omega, \lambda}U^{\dagger}_{\mu\lambda}(\mathbf{k})E_{\lambda}(\mathbf{k},\omega)
e^{i\mathbf{k}\cdot \left( \mathbf{r} +\mathbf{e}_{\mu}/2\right) - i\omega \tau },
\nonumber\\
\end{eqnarray}
where $\lambda=1,2,3,4$ index the normal modes of the action, the wavevectors $\mathbf{k}$ are summed over the first Brillouin zone of the fcc lattice, $\omega$ are Matsubara frequencies, $N_s$ is the number of sites of the fcc lattice, and $U^{\dagger}_{\mu\lambda}(\mathbf{k})$ are $4\times 4$ unitary matrices. In this basis the ${\rm curl}$ operator can be written as

\begin{eqnarray}
{\rm curl}_{ij} (A)
&=&
\frac{1}{\sqrt{N_s \beta}}
\sum_{\mathbf{k} \in {\rm BZ}}\sum_{\mu,\lambda} 
Z_{\mu\nu}(\mathbf{k})U^{\dagger}_{\nu\lambda}(\mathbf{k})
\nonumber\\
&\times&
e^{i\mathbf{k}\cdot(\mathbf{r}_{\alpha}+\mathbf{e}_{\mu}/2) -i \omega \tau}
A_{\lambda}(\mathbf{k}, \omega)
\nonumber\\
&=&
\frac{1}{\sqrt{N_s \beta}}
\sum_{\lambda,\mathbf{k} \in {\rm BZ}}
\xi_{\lambda}(\mathbf{k})U^{\dagger}_{\mu\lambda}(\mathbf{k})
\nonumber\\
&\times&
e^{i\mathbf{k}\cdot(\mathbf{r}_{i}+\mathbf{e}_{\mu}/2) -i \omega \tau}
A_{\lambda}(\mathbf{k}, \omega),
\nonumber\\ \label{eq: diagonal curl}
\end{eqnarray}
where the columns of the matrix $U^{\dagger}_{\nu\lambda}(\mathbf{k})$ are eigenvectors of the matrix $Z_{\mu \nu}(\mathbf{k})=2i\sin\left(\mathbf{k}\cdot\mathbf{\Delta}_{\mu \nu}\right)$ with eigenvalues
\begin{eqnarray}
&&\xi_{\lambda=1,2}( \mathbf{k})
=
\pm\sqrt{2}\sqrt{\sum_{\mu \nu}\sin^2 \left( \mathbf{k} \cdot  \mathbf{\Delta}_{\mu \nu}\right)}
\, ,
\nonumber\\
&&
\xi_{\lambda =3,4}( \mathbf{k}) 
=0 
\, .
\label{eq: xi}
\end{eqnarray}
This identifies $\lambda=3,4$ as the longitudinal modes since they vanish under the action of the ${\rm curl}$ operator.
It follows that the transverse part of the action becomes diagonal in the new basis
\begin{eqnarray}
&&\mathcal{S}_{\rm tran.}=\sum_{\omega}\sum_{\mathbf{k}\in{\rm BZ},\lambda=1,2}
\Big[\omega A_{\lambda}(-\mathbf{k},-\omega)E_{\lambda}(\mathbf{k},\omega)
\nonumber\\
&&+ \tilde{g}E_{\lambda}(-\mathbf{k},-\omega) E_{\lambda}(\mathbf{k},\omega) 
+ \frac{z\tilde{g}}{s^2}  \xi_{\lambda}^2(\mathbf{k})  A_{\lambda}(-\mathbf{k},-\omega)   A_{\lambda}(\mathbf{k},\omega)     
\nonumber\\
&&+2\pi i    j_{\lambda}(\mathbf{k},\omega) A_{\lambda}(-\mathbf{k},-\omega)  \Big] ,
\label{eq: transverse}
\end{eqnarray}
where in the the Coulomb gauge $A_{\lambda=3,4}(\mathbf{k}, \omega )=0$ and
\begin{eqnarray}
 j_{\lambda}(\mathbf{k},\omega) &=&\frac{1}{\sqrt{N_s \beta}}
\sum_{\mu, \tau, \mathbf{r} \in {\rm fcc}} U _{\lambda \mu}   (\mathbf{k})
 j_{\mu}(\mathbf{r}+\mathbf{e}_{\mu}/2,\tau)
\nonumber\\
&&
\times
e^{i\omega \tau - i\mathbf{k} \cdot \left(  \mathbf{r} + \mathbf{e}_{\mu}/2  \right)}.
\end{eqnarray}

\subsubsection{Longitudinal modes}

We proceed to finding the normal modes of the longitudinal part of the action. The constraint in Eq.~\eqref{eq: B quantisation}  uniquely determines the longitudinal ($\lambda=3,4$) modes of the electric field $E_{\alpha\beta}(\tau)$ via the lattice Laplace equation. To form the equation, we write the longitudinal field as the lattice derivative 
\begin{eqnarray}
E(\mathbf{r} + \mathbf{e}_{\mu}/2, \tau) = \Phi_{\rm d} (\mathbf{r} + \mathbf{e}_{\mu}, \tau) - \Phi_{\rm u} (\mathbf{r}, \tau ),
\end{eqnarray}
where the variables $\Phi_{\rm u/d}(\mathbf{r})$ are defined on the centres of up/down tetrahedra that make up two fcc lattices. Making the above substitution, the constraint in Eq.~\ref{eq: B quantisation}, for the up tetrahedra,  can be written as
\begin{eqnarray}
2\pi q_{\rm u} (\mathbf{r}, \tau) = \sum_{\mu} \left[  \Phi_{\rm d} (\mathbf{r} + \mathbf{e}_{\mu}, \tau)
-\Phi_{\rm u} (\mathbf{r}, \tau)
 \right],
\end{eqnarray}
and for the down tetrahedra as 
\begin{eqnarray}
2\pi q_{\rm d} (\mathbf{r}, \tau) = -\sum_{\mu} \left[  \Phi_{\rm d} (\mathbf{r}, \tau)
-\Phi_{\rm u} (\mathbf{r} - \mathbf{e}_{\mu}, \tau)
 \right],
\end{eqnarray}
where $q_{\rm u} (\mathbf{r}, \tau)$ and $q_{\rm d} (\mathbf{r}, \tau)$ are the vison occupation numbers at position $\mathbf{r}$ of an up or down tetrahedron respectively. Fourier transforming the two equations we obtain a unique solution (up to a constant) for $\Phi_{\rm u/d }(\mathbf{r}, \tau)$
\begin{eqnarray}
\begin{pmatrix}
\Phi_{\rm u}(\mathbf{k}, \tau) \\ \Phi_{\rm d}(\mathbf{k}, \tau)
\end{pmatrix}
=\frac{\pi}{2\left(  |\gamma(\mathbf{k})|^2 - 1  \right) }
\begin{pmatrix}
1 & \gamma(\mathbf{k})  \\ \gamma^{\ast}(\mathbf{k}) & 1
\end{pmatrix}
\begin{pmatrix}
q_{\rm u}(\mathbf{k}, \tau) \\ q_{\rm d}(\mathbf{k}, \tau)
\end{pmatrix}
,
\nonumber\\
\label{eq: Laplace}
\end{eqnarray}
where $\gamma(\mathbf{k})=\frac{1}{4}\sum_{\mu} e^{i\mathbf{k}\cdot\mathbf{e}_{\mu}}$, $\Phi_{\rm u/d}(\mathbf{k})= \frac{1}{\sqrt{N_s}}\sum_{\mathbf{r} \in {\rm  u/d}} e^{-i \mathbf{k} \cdot \mathbf{r}} \Phi_{\rm u/d}(\mathbf{r})$ and $q_{\rm u/d}(\mathbf{k})= \frac{1}{\sqrt{N_s}}\sum_{\mathbf{r} \in {\rm  u/d}} e^{-i \mathbf{k} \cdot \mathbf{r}} q_{\rm u/d}(\mathbf{r})$ ($\mathbf{r}$ is summed over the positions of up/down tetrahedra respectively).
Substituting this back into $\mathcal{S}$ in Eq.~\ref{partition and action}, we obtain the unique longitudinal part of the action
\begin{eqnarray}
\mathcal{S}_{\rm long.} &=& \int_{-\beta/2}^{\beta/2} d\tau \Big[   \sum_{\mathbf{r} \mathbf{r}' \in {\rm u}}  
V(\mathbf{r} - \mathbf{r}')    q_{\rm u}(\mathbf{r}, \tau) q_{\rm u}(\mathbf{r}', \tau) 
\nonumber\\
&+&
 \sum_{\mathbf{r} \mathbf{r}' \in {\rm d}} V(\mathbf{r} - \mathbf{r}')  q_{\rm d}(\mathbf{r}, \tau) q_{\rm d} (\mathbf{r}', \tau)
\nonumber\\
&+&
\sum_{\mathbf{r} \in {\rm u}} \sum_{\mathbf{r}' \in {\rm d}} V_{\rm ud} (\mathbf{r} - \mathbf{r}')
  q_{\rm u}(\mathbf{r}, \tau) q_{\rm d} (\mathbf{r}', \tau)
\Big],
\end{eqnarray}
where
\begin{eqnarray}
V(\mathbf{r} - \mathbf{r}') &=& \frac{\tilde{g}}{N_s}\sum_{\mathbf{k} \in {\rm BZ}} \frac{\pi^2}{1- |\gamma(\mathbf{k})|^2}
e^{i \mathbf{k} \cdot \left( \mathbf{r} - \mathbf{r}' \right)}
,
\nonumber\\
V_{\rm ud}(\mathbf{r} - \mathbf{r}') &=&  \frac{\tilde{g}}{N_s}\sum_{\mathbf{k} \in {\rm BZ}} 
\frac{2\pi^2 \gamma(\mathbf{k})}{1 - |\gamma(\mathbf{k})|^2}
e^{i \mathbf{k} \cdot \left( \mathbf{r} - \mathbf{r}' \right)}
.
\end{eqnarray}
Assuming the visons are far apart, it is useful to decompose the sum into its diagonal and off-diagonal parts


\begin{eqnarray}
\mathcal{S}_{\rm long.} &=&
\int_{-\beta/2}^{\beta/2} d\tau\: \mu_{\rm V} \sum_{\sigma=u/d,\mathbf{r} \in {\rm fcc}} q^2_{u}(\mathbf{r},\tau) 
 \nonumber\\
 &&
 +
 \int_{-\beta/2}^{\beta/2} d\tau \sum_{\sigma,\sigma',\mathbf{r}, \mathbf{r}' \in {\rm fcc}}
\tilde{V} (\mathbf{r} - \mathbf{r}')
 q_{\sigma} (\mathbf{r}, \tau) q_{\sigma'} (\mathbf{r}', \tau),
 \nonumber\\
\end{eqnarray}
 where $\tilde{V} (\mathbf{r})$ is the asymptotic Coulomb part of the vison interaction energy with the additional constraint that $\tilde{V} (\mathbf{0})=0$
and the vison chemical potential (self-energy) is given by
\begin{eqnarray}
\mu_V = \frac{\tilde{g}}{N_s}\sum_{\mathbf{k} \in {\rm BZ}} 
\frac{\pi^2 }{1 - |\gamma(\mathbf{k})|^2} =C_1\tilde{g},
\end{eqnarray}
where $C_1=17.5$. For $\mathbf{r}\neq \mathbf{0}$:
\begin{eqnarray}
\tilde{V}(\mathbf{r})= \int \frac{d^3 \mathbf{k}}{(2\pi)^3} \frac{8\tilde{g}a_0\pi^2}{|\mathbf{k}|^2}e^{i \mathbf{k}\cdot\mathbf{r}}.
\label{eq: vison Coulomb}
\end{eqnarray}
The above vison self-energy and Coulomb interaction agree with the results of Ref.~\onlinecite{Attila}.

\section{Instantons} \label{Instantons}

\subsection{Stationary solutions of the action}

Stationary solutions of the Euclidean action make an important non-perturbative ($\propto e^{-s}$) contribution to the partition function, that restores the periodic symmetry with respect to the electric field $E_{\alpha\beta}(\tau)$. The instanton is precisely such a stationary solution. For illustration, let us first consider a single instanton at  $\mathbf{r}=\mathbf{0}$ and imaginary time $\tau=0$ in the direction $\mathbf{e}_{\sigma}$, i.e.,  $j_{\mu=\sigma}(\mathbf{0}, 0)=1$ with all other $j_{\alpha\beta}$ vanishing. To satisfy the continuity equation in Eq.~\ref{continuity}, we will consider a vison of charge $2\pi$ tunneling from the up tetrahedron at $\mathbf{r}=-\mathbf{e}_{\sigma}/2$ to the down tetrahedron at $\mathbf{r}=\mathbf{e}_{\sigma}/2$. Correspondingly, the vison occupation number $q_{\rm u}(-\mathbf{e}_{\sigma}/2, \tau)$ $\left ( q_{\rm d}(\mathbf{e}_{\sigma}/2,\tau) \right)$ decreases (increases) at $\tau=0$ by one. (Alternatively, we could have considered a vison of charge $-2\pi$ tunneling in the opposite direction or a creation of a vison-antivison pair.)

When deriving the stationary solution of the action, it proves convenient to briefly reinstate the longitudinal part of $A_{\alpha\beta} (\tau)$ ($A(\mathbf{k}, \omega)_{\lambda=3,4}$ normal modes) in the action so that the continuity expressed in Eq.~\ref{continuity} is automatically taken care of. (This, of course, has no effect on the physical, gauge-invariant observables and is the only place where we are not working in the Coulomb gauge). We can see how continuity is enforced as follows. Substituting the longitudinal part of $A_{\alpha\beta} (\tau)$, which can be written as $A_{\alpha\beta} (\tau)=\chi_{\beta}(\tau) - \chi_{\alpha}(\tau)$, into the action $\mathcal{S}$, and collecting all terms where it enters, we obtain
\begin{eqnarray}
i\sum_{\alpha,\tau} \chi_{\alpha} (\tau) \left[  {\rm div}_{\alpha} \vartriangle_{\tau}E(\tau) 
- 2\pi {\rm div}_{\alpha} j(\tau) \right].
\end{eqnarray}
The longitudinal part $\chi_{\alpha} (\tau)$ can thus be interpreted as a Lagrange multiplier that enforces the continuity expressed in Eq.~\ref{continuity}. The instanton solution can now be simply derived from the transverse part of the action in Eq.~\ref{eq: transverse} by extending the sum to include the longitudinal modes $\lambda = 3,4$. Minimising $\mathcal{S}$ in Eq.~\ref{eq: transverse} with respect to variations in the variables $E_{\lambda} (\mathbf{k}, \omega)$ and $A_{\lambda} (\mathbf{k}, \omega)$, we obtain the stationary solution of the action
\begin{eqnarray}
&&\omega A_{\lambda}(\mathbf{k},\omega)=2\tilde{g}E_{\lambda}(\mathbf{k},\omega)
\nonumber\\
&&\omega E_{\lambda}(\mathbf{k}, \omega)
+\frac{2z\tilde{g}}{s^2}\xi_{\lambda}^2(\mathbf{k})A_{\lambda}(\mathbf{k}, \omega)
+\frac{2\pi i}{\sqrt{N_s \beta}} U_{\lambda \sigma}(\mathbf{k})=0.
\nonumber\\
\end{eqnarray}
Solving for the electric field, we obtain
\begin{eqnarray}
E^{\rm inst.}_{\lambda}(\mathbf{k},\omega)=E^{\rm const.}_{\lambda}(\mathbf{k})\delta_{\omega,0}
-\frac{1}{\sqrt{N_s \beta}}
\frac{\pi i \omega U_{\lambda \sigma}(\mathbf{k})}{\frac{2z\tilde{g}^2}{s^2}\xi_{\lambda}^2(\mathbf{k})
+\frac{\omega^2}{2}},
\nonumber\\
\label{eq: instanton solution}
\end{eqnarray}
where $E^{\rm const.}_{\lambda}(\mathbf{k})$ is constant in time and is a purely divergenceful field (i.e. $E^{\rm const.}_{\lambda=1,2}(\mathbf{k})=0$) due to a pair of vison charges $\pi$ at positions $\pm \mathbf{e}_{\mu}/2$. This is to ensure that we are describing the tunneling of vison of charge $2\pi$ at $\tau=0$. (See App.~\ref{app: instanton continuum} for further details.)

It is important to check that our solution lies within the domain $|E_{\alpha\beta} (\tau)| \leq \pi$. By symmetry, the solution will reach its maximum at $\tau=0$, $\mathbf{r}=\mathbf{0}$
\begin{eqnarray}
E_{\sigma}^{\rm inst.} (\mathbf{0}, 0) &=& 
 \int_{-\infty}^{\infty} 
 \frac{d \omega}{2\pi N_s} 
 \sum_{\lambda, \mathbf{k} \in {\rm BZ}} 
\frac{-\pi i \omega U^{\dagger}_{\sigma \lambda}(\mathbf{k})U_{\lambda \sigma}(\mathbf{k})}{\frac{2z\tilde{g}^2}{s^2}\xi_{\lambda}^2(\mathbf{k})
+\frac{\omega^2}{2}}
\nonumber\\
&=&\pm \pi,
\end{eqnarray}
\label{eq: discontinuity}
where the field jumps from $-\pi$ to $\pi$ at $\tau=0$ and we have used the fact that $E^{\rm const.}_{\mu} (\mathbf{0})=0$. We have thus verified that $|E_{\alpha\beta} (\tau)| \leq \pi$ everywhere.    Note that we are working in the low-temperature limit $\beta \tau_{\rm QF} \gg 1$, where finite-size effects in imaginary time can be neglected and $\sum_{\omega \in 2\pi n/\beta} \approx \int_{-\infty}^{\infty} \beta \frac{d \omega}{2\pi}$. $\tau_{\rm QF}=\frac{s}{\tilde{g}\sqrt{z}}$ is the characteristic timescale of quantum fluctuations. Notice that $\beta \tau_{\rm QF} \gg 1$ implies that the the temperatures are low by comparison with the photon bandwidth\cite{Our_Work}.

\subsection{The instanton measure}
To compute the measure associated with a single instanton, we first remove the discontinuity in the instanton solution $E^{\rm inst.}_{\mu} (\mathbf{r}, \tau)$ at $\tau=0$ (see Eq.~\ref{eq: discontinuity}) and let $E_{\mu} (\mathbf{0}, \tau)$ wind by $2\pi$ instead
\begin{eqnarray}
E_{\mu} (\mathbf{r}, \tau) \rightarrow E_{\mu} (\mathbf{r},\tau) 
+ \delta_{\mathbf{r}, \mathbf{0}}   \delta_{\mu\sigma} \left[ \pi{\rm sgn} (\tau) +\pi \right].
\end{eqnarray}
Note that this transformation removes the jump in the longitudinal part of the electric field that occurs at $\tau=0$ (see App.~\ref{app: smooth instanton}). $E^{\rm long.}_{\mu} (\mathbf{r}, \tau)$ is now the electric field due to a vison of charge $2\pi$ at $\mathbf{r} = -\mathbf{e}_{\sigma}/2$ for $\tau >0$ as well as $\tau<0$. The longitudinal part of the electric field is thus a constant and the dynamical part is purely transverse: $\left(  \lambda=1,2  \right)$  are the only components we need to consider, when calculating the instanton measure.
Considering the action $\mathcal{S}$ in Eq.~\ref{partition and action}, the above transformation removes the $2 \pi i j_{\mu} (\mathbf{0}, 0) A_{\mu} (\mathbf{0}, 0)$ term, translates the quadratic potential $E_{\alpha\beta}^2$ by $2\pi$, alters the constraint enforced by $\varphi_{\alpha} (\tau)$ so that ${\rm div}E = 2\pi$ at $\mathbf{r}=\mathbf{e}_{\mu}/2$ for all $\tau$, and changes the range of integration over $E_{\mu} (\mathbf{0}, \tau)$ from $|E_{\mu} (\mathbf{0},\tau)| \leq \pi$ to $\pi{\rm sgn} (\tau) < E_{\mu} (\mathbf{0}) \leq 2\pi + \pi{\rm sgn} (\tau) $ (the range of integration over the other $E_{\alpha\beta} (\tau)$ variables remains unaltered). Integrating out the variables $A_{\lambda=1,2} (\mathbf{k}, \omega)$ in the transverse part of the action $\mathcal{S}_{\rm tran.}$, we are left with the following action for the variables $E_{\lambda=1,2} (\mathbf{k}, \omega)$ together with a { \it smooth} instanton solution $E^{\rm sm.}_{\mu} (\mathbf{r}, \tau) $
\begin{eqnarray}
\mathcal{S}_{\rm tran} \left[   E_{\lambda} (\mathbf{k}, \tau)  \right] 
&=&
\int_{-\beta/2}^{\beta/2} d\tau 
\Big{[}
\sum_{\mathbf{k} \in {\rm BZ}, \lambda=1,2} 
 \frac{s^2  |\dot{E}_{\lambda} (\mathbf{k}, \tau)|^2 }{4z\tilde{g}\xi_{\lambda}^2(\mathbf{k})} 
\nonumber\\
&& + 
\tilde{g}\sum_{\alpha\beta}V \left(E_{\alpha\beta}(\tau)\right)
\Big{]}
, \label{eq: instanton action}
\\
E^{\rm sm.} _{\mu}(\mathbf{r}, \tau) &=&
 E^{\rm inst.}_{\mu} (\mathbf{r}, \tau) 
+ \delta_{\mathbf{r}, \mathbf{0}}   \delta_{\mu\sigma} \left[  \pi{\rm sgn} (\tau) +\pi \right], 
\label{eq: smooth instanton}
\nonumber\\
\end{eqnarray}
where $V \left(E_{\alpha\beta}(\tau)\right) = {\rm min}_n \left(   E_{\alpha\beta}(\tau) - 2\pi n   \right)^2$ is a continued parabolic potential and the minimum is taken with respect to integer $n$. See App.~\ref{app: smooth instanton} for verification that $E^{\rm sm.} _{\mu}(\mathbf{r}, \tau)$ is indeed a stationary solution of the above action.

Fluctuations around the {\it smooth} instanton solution $\delta E_{\mu} (\mathbf{r}, \tau) = E_{\mu} (\mathbf{r}, \tau) - E_{\mu}^{\rm sm.} (\mathbf{r}, \tau) $ are themselves smooth and are governed by the following action
\begin{eqnarray}
\delta \mathcal{S}_{\rm tran} \left[    \delta E_{\lambda} (\mathbf{k}, \tau)  \right] &=&
\int_{-\frac{\beta}{2}}^{\frac{\beta}{2}} d\tau
\sum_{\mathbf{k} \in {\rm BZ}, \lambda=1,2} 
 \frac{ s^2  |\delta \dot{E}_{\lambda} (\mathbf{k}, \tau)|^2  }{4z\tilde{g}\xi_{\lambda}^2(\mathbf{k})} 
\nonumber\\
& +& \tilde{g}\int_{-\frac{\beta}{2}}^{\frac{\beta}{2}} d\tau
\sum_{\alpha\beta} \frac{1}{2} V'' \left(E^{\rm sm.}_{\alpha\beta}(\tau)\right) \delta E^2_{\alpha\beta} (\tau),
\nonumber\\
\label{eq: instanton fluctuations}
\end{eqnarray}
where $V'' \left(E^{\rm sm.}_{\mu}(\mathbf{r}, \tau)\right) = -4\pi \delta(E^{\rm sm.}_{\mu}(\mathbf{r},\tau)-\pi) + 2 = -4\pi \delta(\tau)\delta_{\mathbf{r},\mathbf{0}}\delta_{\mu\sigma}/\dot{E}^{\rm sm.}_{\sigma}(\mathbf{0},0)+2$. The normal modes of this action, which include the zero translational mode $\dot{E}^{\rm sm.}(\mathbf{r},\tau)$ (see App.~\ref{app: smooth instanton} for an explicit proof) determine the instanton measure
\begin{eqnarray}
d\tilde{\xi}_0 \frac{\sqrt{\langle \dot{E}^{\rm sm.} | \dot{E}^{\rm sm.}  \rangle}}{\sqrt{2\pi}} 
\left(\frac{{\rm det}' \tilde{K} }{{\rm det}\; K}\right)^{-\frac{1}{2}}=\frac{gC_3d\tilde{\xi}_0}{\sqrt{s}},
\end{eqnarray}
where $C_3=5.13$, $\tilde{\xi}_0$ specifies the position of the instanton core, ${\rm det}' \tilde{K} $ is the determinant of the kernel of the fluctuation action in Eq.~\ref{eq: instanton fluctuations}, excluding the zero eigenvalue, and ${\rm det}\; K $ is the determinant of the kernel of the original action in Eq.~\ref{eq: instanton action}. See App.~\ref{app: instanton measure} for a detailed derivation of the measure.

\subsection{Instanton interactions}
The power-law decay (see App.~\ref{app: instanton continuum}) of the single instanton solution gives rise to long-range interactions between instantons. This infrared effect has consequences for the low-energy properties of the system. Just like the visons interact via the Coulomb potential, the instantons, which correspond to vison currents, interact in Euclidean spacetime via forces that obey the inverse square law. 

If the instantons are far apart the interactions will be taking place in a region of space with small electric field. To derive the long-range part of instanton interactions, we can therefore relax the constraint on the electric field $E_{\alpha\beta}\in (-\pi,\pi]$ and integrate it out to obtain an effective action $\mathcal{S}_{\rm inst.}$ that describes instanton-instanton interactions, i.e., interactions between vison currents,
\begin{eqnarray}
\mathcal{S}_{\rm inst.} = 
\sum_{\tau, \tau'} \sum_{\mathbf{r}, \mathbf{r}' \in {\rm fcc}}
V_{\mu\nu} (\mathbf{r} - \mathbf{r}', \tau - \tau')
 j_{\mu} (\mathbf{r}, \tau) j_{\nu} (\mathbf{r}', \tau'),
 \nonumber\\
\end{eqnarray}
where
\begin{eqnarray}
&&V_{\mu\nu} (\mathbf{r} - \mathbf{r}', \tau - \tau') = 
\nonumber\\
&&\frac{\pi^2}{N_s} \sum_{\lambda=1,2, \mathbf{k} \in {\rm BZ}}
\int_{-\infty}^{\infty} \frac{d \omega}{2\pi} 
\frac{e^{i \omega (\tau' -\tau) + i \mathbf{k} \cdot \left( \mathbf{r} - \mathbf{r}' \right)}}
{\frac{\omega^2}{4\tilde{g}} + \frac{z\tilde{g}}{s^2} \xi_{\lambda}^2(\mathbf{k})}
\nonumber\\
&&\times
U^{\dagger}_{\mu\lambda} (\mathbf{k}) U_{\lambda \nu} (\mathbf{k}).
\end{eqnarray}
In the limit where instantons are far apart, it is useful to decompose the sum into its diagonal and off-diagonal parts
\begin{eqnarray}
\mathcal{S}_{\rm inst.} &=& \sum_{\tau, \tau'} \sum_{\mathbf{r}, \mathbf{r}' \in {\rm fcc}}
\tilde{V}_{\mu\nu} (\mathbf{r} - \mathbf{r}', \tau - \tau')
 j_{\mu} (\mathbf{r}, \tau) j_{\nu} (\mathbf{r}', \tau')
 \nonumber\\
 &&
 + \mu_{\rm I} \sum_{\tau, \mathbf{r} \in {\rm fcc}} j^2_{\mu}(\mathbf{r},\tau) ,
\end{eqnarray}
where $\tilde{V}_{\mu\nu} (\mathbf{r} - \mathbf{r}',\tau - \tau')$ is given by the asymptotic inverse square law of $V_{\mu\nu} (\mathbf{r} - \mathbf{r}',\tau - \tau')$ with the additional constraint that $\tilde{V}_{\mu\mu}(\mathbf{0},0)=0$, and
the instanton chemical potential (its self-energy) is given by
\begin{eqnarray}
\mu_{\rm I} =V_{11} (\mathbf{0},0) &=&  \frac{\pi^2}{N_s} \sum_{\lambda=1,2, \mathbf{k} \in {\rm BZ}} \int_{-\infty}^{\infty} \frac{d \omega}{2\pi} 
\frac{U^{\dagger}_{\mu\lambda} (\mathbf{k}) U_{\lambda \nu} (\mathbf{k})}
{\frac{\omega^2}{4\tilde{g}} + \frac{z\tilde{g}}{s^2} \xi_{\lambda}^2(\mathbf{k})} \nonumber\\
&=&
\frac{\pi^2s}{\sqrt{z}N_s}
\sum_{\mathbf{k} \in {\rm BZ}}
\frac{Z_{11}^2 (\mathbf{k})}{|\xi_1(\mathbf{k})|^3} =C_2s,
\end{eqnarray}
where $Z^2_{11}(\mathbf{k})$ is given by the matrix product $\sum_{\mu} Z_{1\mu}(\mathbf{k})Z_{\mu 1}(\mathbf{k})$ and $C_2=0.624$. (Note that the self-energy is exact, i.e. unaffected by the above relaxation of the constraint on the size of the electric field and can be obtained by substituting the single instanton solution in Eq.~\ref{eq: instanton solution} into the action.) The above decomposition separates the UV (self-energy) and IR (long-range interaction) contributions to the instanton action and can equivalently be obtained by modifying the original action in Eq.~\ref{Action} as follows. One imposes a small cutoff $\Lambda << a_0$ on the action, i.e. any fluctuations of the gauge fields whose wavelength is not much greater than the lattice spacing are quenched. This has no effect on long-range photon correlations or interactions between instantons or visons that are far apart. It does however give a vanishingly small instanton and vison self-energies, i.e. $\mu_I\sim(\Lambda/a_0)^2\ll 1$ and $\mu_V\sim (\Lambda/a_0)\ll 1$, and so the instanton and vison chemical potential terms need to be added to the action in Eq.~\ref{Action} to compensate for this. We thus obtain an effective model with a reduced cutoff described by the following action
\begin{eqnarray}
&&\mathcal{S}= \mathcal{S}_0 +\int_{-\beta/2}^{\beta/2} d\tau \sum_{\gamma}\Big[ \mu_V q_{\gamma}(\tau)^2 +2\pi i q_{\gamma}(\tau)\phi_{\gamma}(\tau)\Big] \nonumber\\
&&+\sum_{\tau,\alpha\beta}\left(\mu_{\rm I}j_{\alpha\beta}(\tau)^2+2\pi i j_{\alpha\beta}(\tau) A_{\alpha\beta}(\tau)\right)
\nonumber\\
&&+\sum_{\tau, \gamma} i\theta_{\gamma}(\tau)\left(\vartriangle_{\tau}q_{\gamma}(\tau)-{\rm div}_{\gamma}j(\tau)\right), 
\label{eq: reduced cutoff action}
\end{eqnarray}
where
\begin{eqnarray}
\mathcal{S}_0 
&=& 
\sum_{\omega, |\mathbf{k}|<\Lambda , \lambda=1,2}\left(\frac{\tilde{g}z|\mathbf{k}|^2a_0^2}{s^2}+\frac{\omega^2}{4\tilde{g}}\right) 
\left| A_{\lambda}(\mathbf{k},\omega) \right|^2 
\nonumber\\
&&+\frac{1}{4\tilde{g}a_0}\sum_{\omega} \int_{|\mathbf{k}|<\Lambda} 
 \frac{d^3 \mathbf{k}}{(2\pi)^3} |\mathbf{k}|^2 |\varphi( \mathbf{k},\omega) |^2,
\end{eqnarray}
and the compact Higgs field $\theta_{\gamma}(\tau) \in (-\pi,\pi]$ was introduced to enforce continuity at each tetrahedron site. $\varphi(\mathbf{k},\tau)$ is the Fourier transform of the coarse-grained scalar potential $\varphi_{\gamma}(\tau)$.

\section{Ground state}\label{Ground_State}

At zero temperature vison-antivison pairs will be short-lived {\it virtual} excitations. Because of the vison continuity conditions they will necessarily be accompanied by instantons forming a loop in Euclidean spacetime. The Boltzmann weight of the shortest loop will scale as $e^{-2C_2 s}$, and so, in the large $s$ limit, the loops will give a small non-perturbative renormalisation of the speed of light. Note that our calculation of this effect would not be numerically accurate for the smallest loops, because the decomposition into UV and IR parts described in the previous section only works if the instantons are far apart in Euclidean spacetime. At low non-zero temperatures, we will be considering {\it free} visons, whose instanton currents will be far apart in the large-$s$ limit. (Visons can live for all imaginary time at non-zero temperature and there is no need to form a closed loop with instantons and another vison.)

We now sum over the $j_{\alpha\beta}(\tau)=\pm 1$ instanton configurations in the action in Eq.~\ref{eq: reduced cutoff action} (which dominate for $s\gg1$) to obtain an O(2) rotor description of visons coupled to the electromagnetic field 
\begin{eqnarray}
&&Z= \int d\theta_{\gamma}(\tau)\; d\varphi_{\gamma}(\tau)\; dA_{\alpha\beta}(\tau)\; e^{-S_0}\sum_n \frac{e^{-n\mu_I }}{n!}\times 
\nonumber\\
&&\left(  \sum_{\alpha\beta}  
\int_{-\frac{\beta}{2}}^{\frac{\beta}{2}} \frac{C_3\tilde{g} \: d \tau}{\sqrt{s}} 
\left(e^{2\pi i A_{\alpha\beta} -i\vartriangle_{\alpha\beta}\theta}+e^{-2\pi i A_{\alpha\beta}+i\vartriangle_{\alpha\beta}\theta}
\right) \right)^n 
\nonumber\\
&&
\equiv e^{-S_{\rm VP}},
\end{eqnarray}
where the vison plasma action is given by
\begin{eqnarray}
&&\mathcal{S}_{\rm VP}=\mathcal{S}_0
\nonumber\\
&&+\sum_{\gamma}\int_{-\frac{\beta}{2}}^{\frac{\beta}{2}} d \tau \left({\mu_V q_{\gamma}(\tau)}^2+2\pi i q_{\gamma}(\tau)\varphi_{\gamma}(\tau) 
-iq_{\gamma}(\tau)\dot{\theta}_{\gamma}(\tau) \right)
\nonumber\\
&&-\int_{-\frac{\beta}{2}}^{\frac{\beta}{2}} \frac{C_3\tilde{g} e^{-\mu_I}\: d \tau}{\sqrt{s}}  \;\sum_{\alpha\beta}\cos\left(\vartriangle_{\alpha\beta}\theta(\tau)-2\pi A_{\alpha\beta}(\tau)\right).\nonumber\\
\end{eqnarray}
This is one of the central results of this paper and an effective model for the original action in Eq.~\ref{Action} derived in the large-$s$ and low-temperature ($\beta\tilde{g}\sqrt{z}/s \gg 1$) limits.

Motivated by the success of our previous work~\cite{Our_Work}, where
a large-s expansion was used to obtain the low-energy spectrum for $s=1/2$, we believe that $s\rightarrow \infty$ under RG and should be treated as a large parameter. Hence, the last term in the above action should always be thought of as a small perturbation ($\tilde{g}e^{-\mu_I}/\sqrt{s}\ll \mu_V$) and does not drive a condensation of visons (definite phase of $\theta_{\gamma}(\tau)$) even down to $s=1/2$, i.e. QSI is in the deconfined phase for all $s$ values.

\section{Non-zero temperatures}
\label{Non-zero_Temperatures}

\subsection{3D plasma of visons}
At non-zero temperatures the visons are not so innocuous. They become {\it real} thermal excitations and form a 3D Coulomb gas. The gas is always in the plasma phase~\cite{Kosterlitz_1977}, however large the vison chemical potential $\mu_V$. To see this, we consider the RG flow equations of the gas parameters.
In the $s\rightarrow\infty$ limit, where the visons are static, and at low temperatures, where $\beta\mu_V \gg 1$, the RG equations for $\beta\mu_V \ll 1$ are given by\cite{Kosterlitz_1977}
\begin{eqnarray}
\frac{dK^{-1}}{dl}&=&K^{-1} + y^2,
\nonumber\\
\frac{dy}{dl}&=&3y - Ky,
\end{eqnarray}
where the fugacity $y\equiv e^{-\beta \mu_V}$ and $K$ is the coefficient of the vison Coulomb interaction energy $\tilde{V}(\mathbf{r})$ in Eq.~\ref{eq: vison Coulomb} multiplied by $\beta$. Under RG, $\beta\mu_V$ grows linearly from its initial bare value of $\sim\beta\tilde{g}$, whereas
$K$ decays exponentially from its initial bare value of $\sim\beta\tilde{g}$. Eventually, once $K$ has decayed sufficiently, the vison self-energy begins to grow linearly. Hence, the visons dissociate before they begin to proliferate. Even though initially the vison self-energy and interaction energy are comparable, at long lengthscales, the interaction can be neglected by comparison with the self-energy and the visons can be treated as free particles. Quantum Monte Carlo calculations of Ref.~\onlinecite{Attila} found that the energy of a nearest-neighbour vison pair is actually lower than that of a single vison, so that at low temperatures vison pairs are the dominant species and the visons form a weak electrolyte. Here, we argue that the model can be mapped, by integrating out tightly bound vison pairs, to a {\it coarse-grained} 3D Coulomb gas where free visons are the dominant species.  Note that a finite vison inertia (finite $s$) should only increase their ability to screen each other, reducing interactions even further.

\subsection{Semiclassical limit at low temperatures}
At very low temperatures the typical vison spacing will be large and their hardcore interactions negligible. We can therefore treat visons as free bosons of two flavours coupled to the electromagnetic field. Introducing two complex bosonic fields $\Psi^{+}_{\gamma}(\tau)$ and $\Psi^{-}_{\gamma}(\tau)$ corresponding to visons and antivisons respectively, the vison plasma action can be written as follows 
\begin{eqnarray}
&&\mathcal{S}_{\rm VP} =
\nonumber\\
&&\mathcal{S}_0 +\sum_{\sigma=\pm}\int_{-\frac{\beta}{2}}^{\frac{\beta}{2}} d\tau \;\Big{[} \sum_{\gamma} \bar{ \Psi }^{\sigma}_{\gamma}\left(
i \partial_{\tau}  
+2\pi i \sigma \varphi_{\gamma} +\mu_V \right) \Psi^{\sigma}_{\gamma}
\nonumber\\
 &&
-\frac{ C_3\tilde{g}e^{-\mu_{ \rm I } } }{\sqrt{s}}\sum_{\alpha\beta} \left(
\bar{ \Psi }^{\sigma}_{\alpha} \Psi^{\sigma}_{\beta} e^{-2 \pi i \sigma A_{\alpha\beta}} +{\rm h.c}
\right)
\Big{]},
\end{eqnarray}
which is the action of lattice bosons coupled to a gauge field. At sufficiently low temperatures, where $\beta\tilde{g} e^{-\mu_{ \rm I }}/\sqrt{s}\gg 1$\footnote{In this limit, the typical instanton separation along $\tau$ will be small by comparison with $\beta$, so that finite-size effects associated with finite temperature are negligible.}, the visons move at the bottom of the band and a continuum (low $\mathbf{k}$) approximation for the bosonic fields can be made
\begin{eqnarray}
\mathcal{S}_{\rm VP}[\Psi(\mathbf{r})] &=& \mathcal{S}_0 
\nonumber\\
&&+\sum_{\sigma=\pm}
\int_{-\frac{\beta}{2}}^{\frac{\beta}{2}} d\tau\int  \frac{d^3\mathbf{r}}{(a_0^3/8)} \Big{[}
\mu_V   |\Psi^{\sigma}|^2 
\nonumber\\
 && + 
\frac{ 8\tilde{g}C_3a_0^2e^{-\mu_{ \rm I } } }{\sqrt{s}} |\nabla\Psi^{\sigma} -2 \pi i \sigma \mathbf{A}\Psi^{\sigma}|^2
\nonumber\\
&&
+i\bar{\Psi^{\sigma}} \left(\partial_{\tau} + 2\pi \sigma \varphi\right)\Psi^{\sigma}
\Big{]},
\end{eqnarray}
where $A^{\mu}(\mathbf{r})=\mathbf{A}(\mathbf{r})\cdot\mathbf{e}_{\mu}$ and $\varphi(\mathbf{r})$ are coarse-grained vector and scalar potentials.
Note that we have neglected non-perturbative corrections to the vison energy $\mu_V$ of order $e^{-C_2 s}$. This is for self-consistency as we have already neglected perturbative corrections which come from higher order terms in the expansion of the Hamiltonian in Eq.~\ref{eq: expanded Hamiltonian}.
The vison plasma action describes bosons with a thermal activation energy of  $\mu_V$, effective mass of $m^{\ast}=\frac{\hbar^2\sqrt{s}e^{\mu_I}}{16C_3a_0^2\tilde{g}}$ and an effective charge of $2\pi \hbar$.

The bosonic excitations can be treated semiclassically at low temperatures because their typical separation $\sim a_0 e^{\beta \mu_V} >> \lambda_{\rm B}$, where $\lambda_{\rm B} = \sqrt{\frac{2\pi \hbar^2 }{kT m^{\ast}}}$ is the thermal de Broglie wavelength of the bosons. The vison excitations are thus not quantum degenerate and can be treated through a single-particle formalism. Considering the Feynman path integral of a single vison excitation, we obtain
\begin{eqnarray}
S_{\rm single} &=& \int_{-\frac{\beta}{2}}^{\frac{\beta}{2}} d\tau \Big{(}\frac{m^{\ast}}{2\hbar^2}\left( \partial_{\tau}\mathbf{x}\right)^2 \pm 2\pi i \mathbf{A}(\mathbf{x},\tau)\cdot\partial_{\tau}\mathbf{x}
\nonumber\\
&&\pm 2\pi i \varphi( \mathbf{x},\tau)
\Big{)},
\end{eqnarray}
where $\mathbf{x}(\tau)$ is the vison's position vector and $\pm$ depends on whether the vison is positively or negatively charged.
The typical quantum-fluctuations in the position of the particle will be
\begin{eqnarray}
\langle|\Delta \mathbf{x}|\rangle \sim \sqrt{2\pi}\beta\langle | \partial_{\tau}\mathbf{x}|\rangle=\sqrt{\frac{2\pi\hbar^2\beta}{m^{\ast}}}\equiv\lambda_B.
\end{eqnarray}
In the semi-classical limit (at low temperatures), fractional variations in $\mathbf{A}( \mathbf{x},\tau)$, or $\varphi( \mathbf{x},\tau)$, due to the fluctuating position of the particle $\mathbf{x}$ will be negligible
\begin{eqnarray}
\frac{|\nabla \mathbf{A}|}{|\mathbf{A}|} \lambda_B \sim \lambda_B\frac{1/R^3}{1/R^2} = \lambda_B/R\ll 1,
\end{eqnarray}
where $R$ is the distance to the nearest vison current. We can therefore approximate
\begin{eqnarray}
\mathbf{A}(\mathbf{x},\tau)&\approx&
\mathbf{A}\left( \langle\mathbf{x}\rangle,\tau\right),
\nonumber\\
\varphi( \mathbf{x},\tau) &\approx& \varphi \left( \langle \mathbf{x} \rangle,\tau \right),
\end{eqnarray}
where $\langle \mathbf{x} \rangle$ is the average position of the vison.
The particle's position can be decomposed as
\begin{eqnarray}
\mathbf{x}(\tau) = \langle \mathbf{x}\rangle + \frac{1}{\sqrt{\beta}}\sum_{\omega \neq 0} \mathbf{x} (\omega) e^{i \omega \tau} ,
\end{eqnarray}
where the Matsubara frequency $\omega = \frac{2\pi n}{\beta}$ and $n$ is an integer. Using the above approximation for the vector and scalar potentials, the single-particle vison action becomes
\begin{eqnarray}
S_{\rm single} &=& \sum_{\omega\neq 0}
\left(\frac{m^{\ast} \omega^2}{2\hbar^2}|\mathbf{x}(\omega)|^2 \pm 2\pi \omega \mathbf{A}\left(\langle\mathbf{x}\rangle, -\omega\right)\cdot  \mathbf{x}(\omega) \right)
\nonumber\\
&\pm& 2\pi i \sqrt{\beta}\varphi\left( \langle \mathbf{x}\rangle, \omega=0\right) 
\end{eqnarray}
We note that the visons do not couple to the static component of the gauge field $\mathbf{A}(\mathbf{x}, \omega=0)$. Therefore the static background field $A^0_{\alpha\beta}$, be it from the half-integer constraint on the magnetic field or from magnetic monopoles introduced into the system, is irrelevant for the dynamical effect that we are describing here.
Integrating over the fluctuations in the particle's position $\mathbf{x}(\omega)$ with $\omega\neq 0$, we obtain
\begin{eqnarray}
S = \frac{2\pi^2\hbar^2}{m^{\ast} }\sum_{\omega \neq 0}  |\mathbf{A}(\langle \mathbf{x}\rangle, \omega)|^2
\pm2\pi i \sqrt{\beta}\varphi\left( \langle \mathbf{x}\rangle, \omega=0\right) .
\nonumber\\
\end{eqnarray}
Summing over many-particle configurations, we obtain the partition function of the system
\begin{eqnarray}
&&Z= \prod_{\gamma,\alpha\beta}\int d\varphi_{\gamma}(\tau)\;dA_{\alpha\beta}(\tau)\;e^{-S_0}
\sum_n \frac{e^{-n\beta \mu_V }}{n!}
 \nonumber\\
 && \times\prod _{i=1}^n \int \frac{d^3\mathbf{x}_i}{(a_0^3/8)}
 \;e^{ -\frac{2\pi^2\hbar^2}{m^{\ast}} \sum_{\omega\neq 0} |\mathbf{A}( \mathbf{x}_i,\omega)|^2}
\nonumber\\
&& \times
\left(e^{2\pi i \sqrt{\beta}\varphi( \mathbf{ x}_i,0)}+e^{-2\pi i\sqrt{\beta}\varphi( \mathbf{x}_i,0)}\right)
\nonumber\\
&&=\prod_{\gamma,\alpha\beta}\int d\varphi_{\gamma}(\tau)\;dA_{\alpha\beta}(\tau)\;
\exp\Big{[}
-S_0 + \int \frac{d^3\mathbf{x}}{(a_0^3/8)}
e^{-\beta \mu_V}
\nonumber\\
&&\times
e^{ -\frac{2\pi^2\hbar^2}{m^{\ast}} \sum_{\omega \neq 0} |\mathbf{A}( \mathbf{x},\omega)|^2 }
2\cos\left(2\pi\sqrt{\beta} \varphi( \mathbf{x},0)\right)\Big{]}.
\end{eqnarray}
The action can be expanded to quadratic order in the Debye limit $e^{-\beta \mu_V} \ll 1$. One can see that the gauge fields are {\it effectively} small in this limit by performing the following rescaling:
\begin{eqnarray}
\mathbf{x} &\rightarrow& e^{\beta \mu_V/3}\mathbf{x},
\nonumber\\
\omega &\rightarrow& e^{-\beta \mu_V/3}\omega,
\nonumber\\
\mathbf{A}(\mathbf{x}, \omega) &\rightarrow& e^{-\beta \mu_V/6} \mathbf{A}(\mathbf{x},\omega)
\nonumber\\
\varphi(\mathbf{x},\omega) &\rightarrow& e^{-\beta\mu_V/6} \varphi( \mathbf{x}, \omega).
\end{eqnarray}
Hence, to quadratic order, the gauge part of the action is given by
\begin{eqnarray}
&&S_{\rm gauge} =
\nonumber\\
&&\sum_{\omega, |\mathbf{k}| <\Lambda, \lambda=1,2} \left(\frac{\tilde{g}z a_0^2|\mathbf{k}|^2}{s^2}+\frac{\omega^2}{4\tilde{g}} +(1-\delta_{\omega,0})\frac{\omega_p^2}{4g}\right) \left| A_{\lambda}(\mathbf{k},\omega) \right|^2 
\nonumber\\
&&+\sum_{\omega} \int_{|\mathbf{k}|<\Lambda}  \frac{d^3 \mathbf{k}}{(2\pi)^3}
\left(
\frac{|\mathbf{k}|^2}{4\tilde{g}a_0}+32\pi^2\beta a_0^{-3} e^{-\beta \mu_V}
\right)|\varphi( \mathbf{k}, \omega) |^2,
\nonumber\\
\label{eq: gapped gauge action}
\end{eqnarray}
where the plasma frequency
\begin{eqnarray}
\omega_p=\sqrt{\frac{128\tilde{g}\pi^2\hbar^2 e^{-\beta \mu_V}}{a_0^2m^{\ast}}},
\end{eqnarray}
agrees with the continuum derivation presented in appendix~\ref{app: continuum}. Note that we have not included tightly-bound vison pairs in the above analysis since they do not contribute to the plasma oscillation of free charges. We have taken the Debye limit, where there are many visons inside the Debye volume. We can extract the Debye length from the above gauge action 
\begin{eqnarray}
\lambda_{B}=a_0 \left(128\pi^2\beta \tilde{g} e^{-\beta \mu_V}\right)^{-\frac{1}{2}},
\end{eqnarray}
and in the Debye limit $e^{\beta \mu_V} \gg 1$, there are indeed many visons $e^{-\beta \mu_V}a_0^{-3} \lambda_{B}^3 \sim e^{\beta \mu_V/2}\gg 1$ in  the Debye volume.
\begin{figure}
\centering
\includegraphics[width=0.9\columnwidth, angle=0]{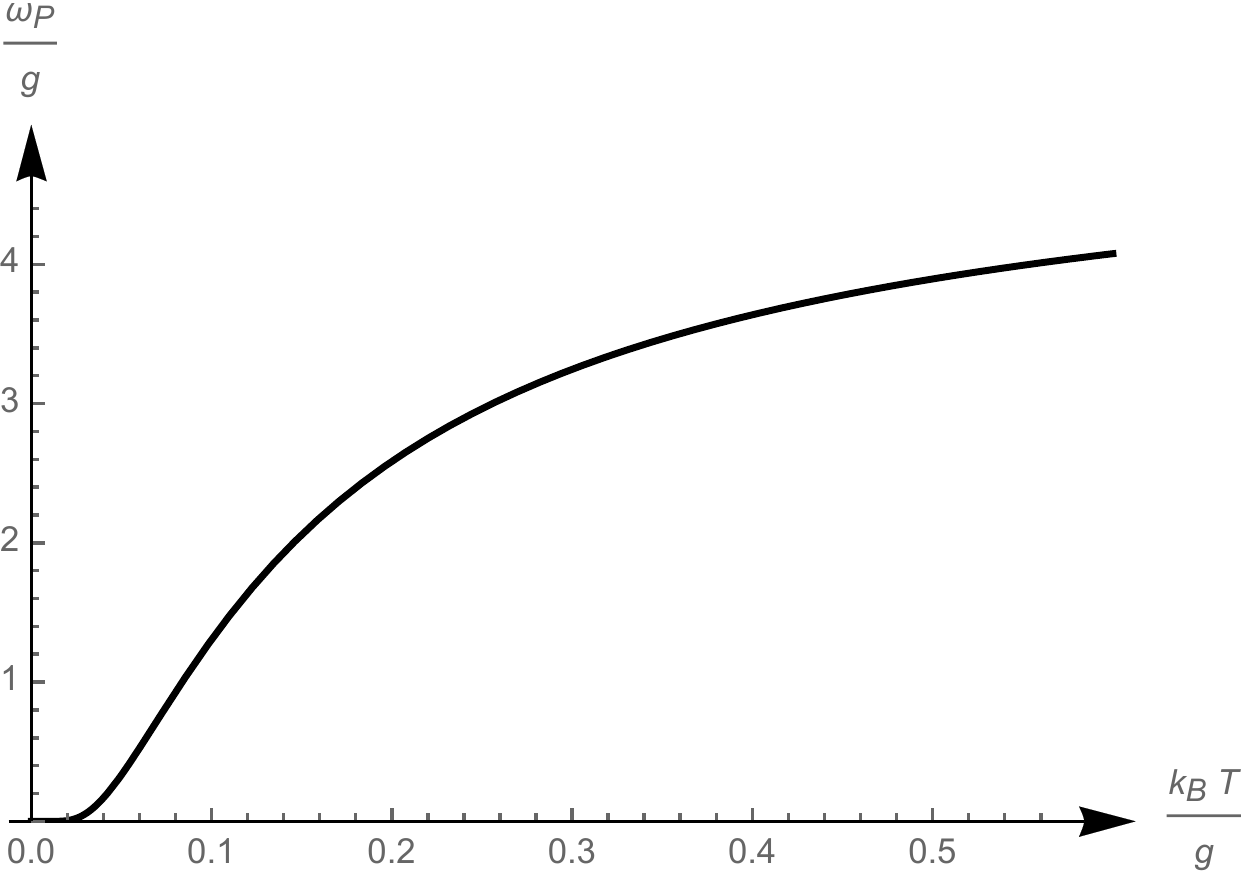}
\caption{
Plasma frequency as a function of temperature.}
\label{FigHexagon}
\end{figure}

\subsection{Correlators and inelastic magnetic response}
We can now extract several correlators from the above quadratic gauge action. In particular, relevant for neutron scattering experiments are magnetic field correlators. For $\omega\neq 0$:
\begin{eqnarray}
\chi(\mathbf{k},i\omega)_{\lambda}\equiv \langle |S^z_{\lambda}(\mathbf{k},\omega)|^2 \rangle
=\frac{4\tilde{g}\xi^2_{\lambda}(\mathbf{k})}{-(i\omega)^2 + E^2_{\mathbf{k}}},
\end{eqnarray}
where $E_{\mathbf{k}}=\sqrt{4\tilde{g}^2z\xi^2_{\lambda}(\mathbf{k})/s^2+\omega_p^2}$. (Notice that we have replaced the low-$\mathbf{k}$ expansion with $\xi_{\lambda}(\mathbf{k})$ to restore periodicity across the Brillouin zone.) For $\omega=0$, $E^2_{\mathbf{k}}$ needs to be replaced with $E^2_{\mathbf{k}}-\omega_p^2$.

\subsubsection{Equal-time structure factor}
\begin{figure}
\centering
\includegraphics[width=0.9\columnwidth, angle=0]{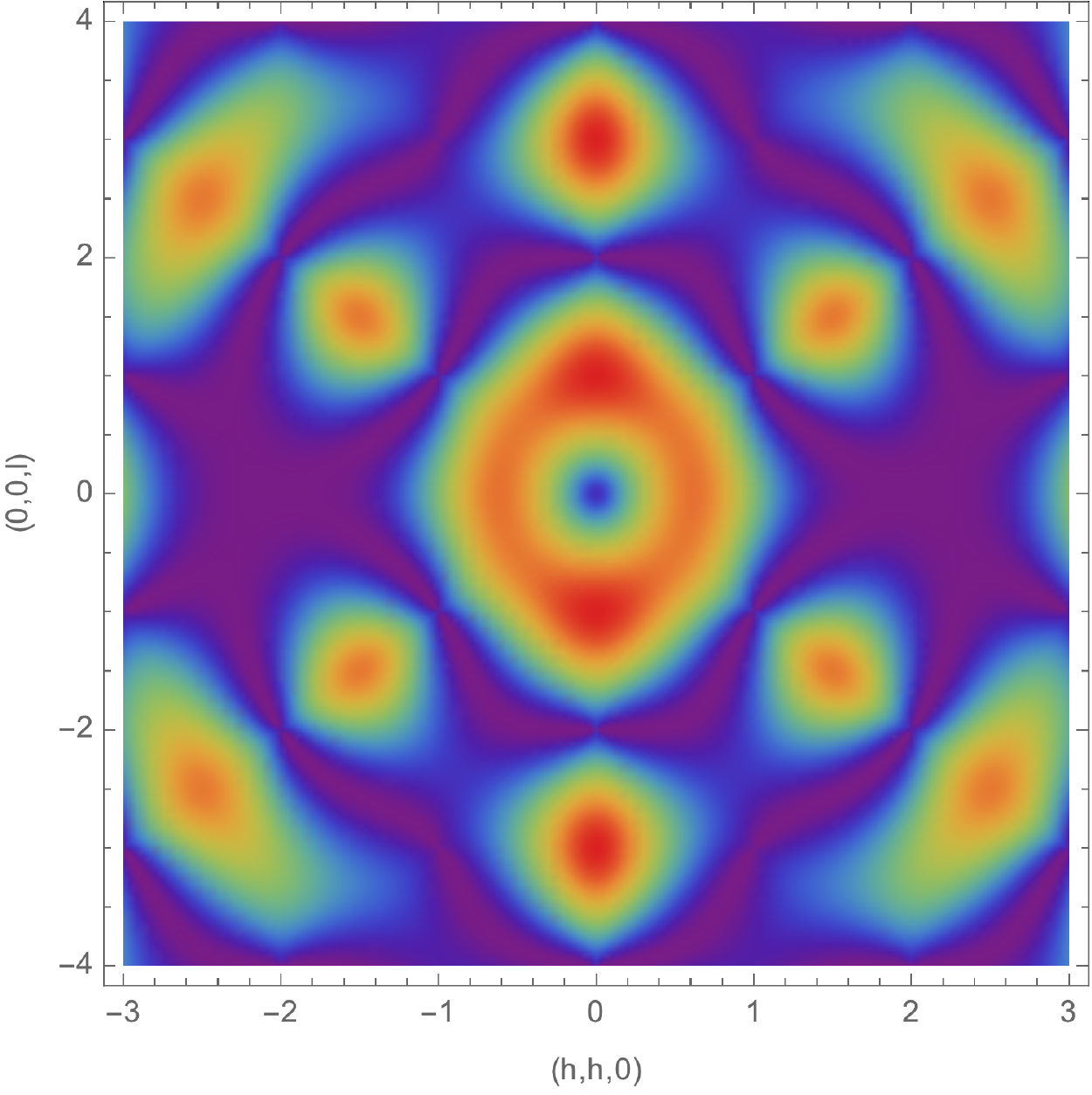}
\caption{
Equal-time structure factor for $s=\frac{1}{2}$ at $T=0.03g$ in the spin-flip channel with $\mathbf{k}=\frac{2\pi}{a_0}(h,h,l)$.}
\label{fig: equal time with visons}
\end{figure}

\begin{figure}
\centering
\includegraphics[width=0.9\columnwidth, angle=0]{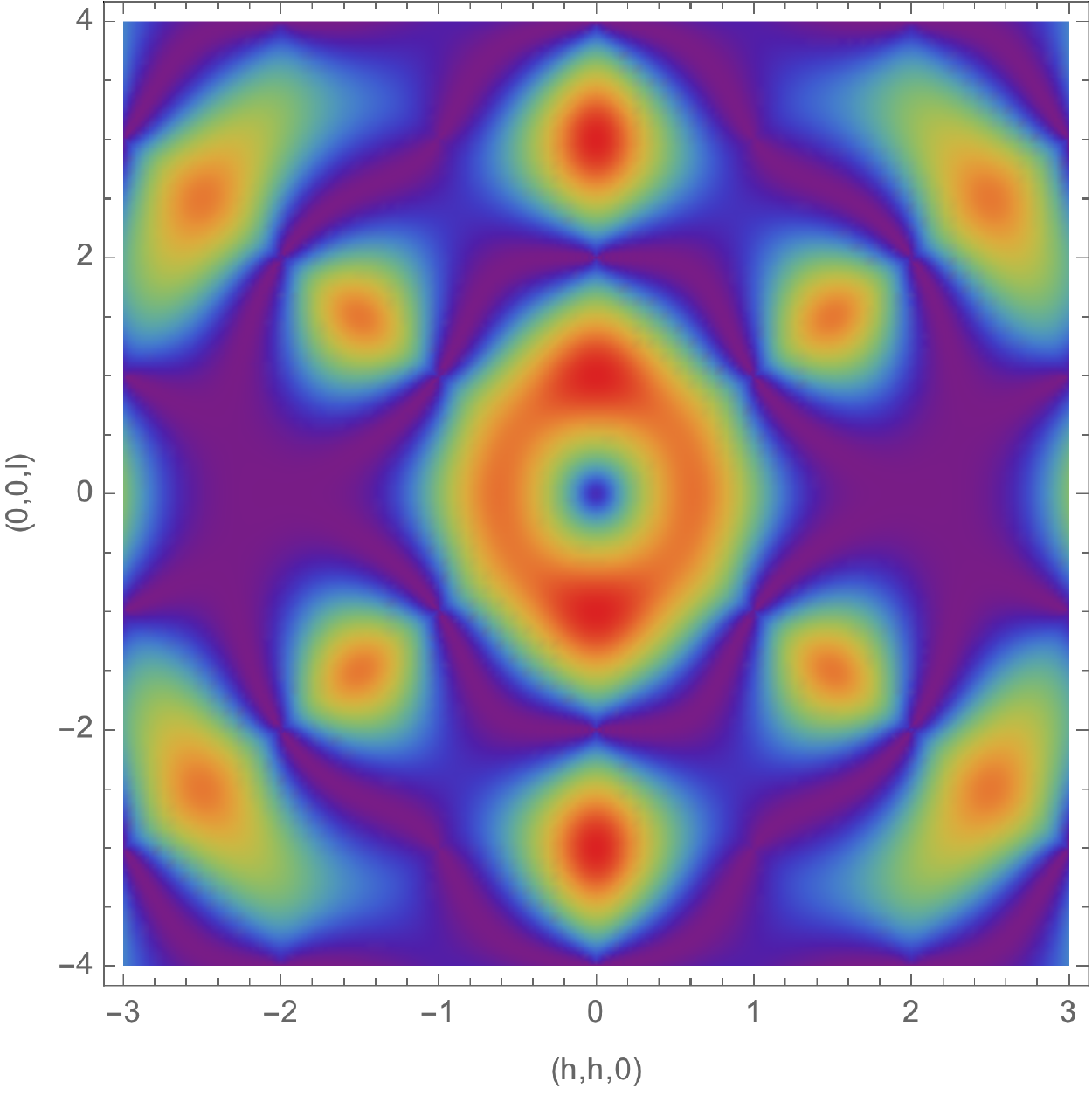}
\caption{
Equal-time structure factor for $s=\frac{1}{2}$ at $T=0.03g$ in the spin-flip channel with $\mathbf{k}=\frac{2\pi}{a_0}(h,h,l)$ without the vison contribution, i.e. with $\omega_p$ set to zero. By comparison with Fig.~\ref{fig: equal time with visons}, we can see that there is a slightly faster restoration of the pinch points without visons.}
\label{fig: equal time with no visons}
\end{figure}
A particularly useful experimental signature in neutron diffraction experiments is the equal-time (energy integrated) structure factor. We calculate the structure factor in the spin-flip channel for a polarised neutron-scattering experiment considered by Ref.~\onlinecite{spin-flip}, 
\begin{eqnarray}
&&S^{yy}(\mathbf{k})\equiv\sum_{\mu ,\nu}
\langle S^z_{\mu}(\mathbf{k},\tau=0) S_{\nu}^z(-\mathbf{k},\tau=0)\rangle
(\hat{\mathbf{e}}_{\mu}\cdot \hat{\mathbf{e}}_y)(\hat{\mathbf{e}}_{\nu}\cdot \hat{\mathbf{e}}_y)
\nonumber\\
&&=
\frac{1}{\beta}\sum_{\mu,\nu}\sum_{\omega,\lambda=1,2} \chi_{\lambda}(\mathbf{k},i\omega) (\hat{\mathbf{e}}_{\mu}\cdot \hat{\mathbf{e}}_y)(\hat{\mathbf{e}}_{\nu}\cdot \hat{\mathbf{e}}_y)
U^{\dagger}_{\mu\lambda}(\mathbf{k})U^{\dagger}_{\nu\lambda}(-\mathbf{k})
\nonumber\\
&&=4\tilde{g}\sum_{\mu\nu}Z^2_{\mu\nu}(\mathbf{k})
(\hat{\mathbf{e}}_{\mu}\cdot \hat{\mathbf{e}}_y)(\hat{\mathbf{e}}_{\nu}\cdot \hat{\mathbf{e}}_y)
\nonumber\\
&&\times
\left(
\frac{1+2n_B(E_{\mathbf{k}}) }{2E_{\mathbf{k}} }+\frac{T}{E^2_{\mathbf{k}}-\omega_p^2}+\frac{T}{E^2_{\mathbf{k}}}
\right),
\end{eqnarray}
where $\mathbf{\hat{e}}_y=\frac{\mathbf{k}\times (1,-1,0)}{|\mathbf{k}\times (1,-1,0)|}$, $n_B(E_{\mathbf{k}})=\frac{1}{e^{\beta E_{\mathbf{k}} } - 1}$ and $Z^2_{\mu\nu}(\mathbf{k})$ is given by the matrix product $\sum_{\sigma} Z_{\mu\sigma}(\mathbf{k})Z_{\sigma\nu}(\mathbf{k})$.

Because the mass gap $\omega_p$ does not couple to the zero Matsubara frequency component, it can only have an effect on the equal-time structure factor at low temperatures, but this is precisely where it is exponentially damped. The effects of the mass gap on the equal-time structure factor are therefore small across the temperature range and cannot be easily seen as shown by Fig.~\ref{fig: equal time with visons} and Fig.~\ref{fig: equal time with no visons}. Motivated by the success of our previous work\cite{Our_Work}, we have extrapolated the results for $s=\frac{1}{2}$. Both figures are in good agreement with the results of Monte Carlo simulations of Ref.~\onlinecite{Shannon}.

\subsubsection{The dynamical structure factor}

\begin{figure}
\centering
\includegraphics[width=0.9\columnwidth, angle=0]{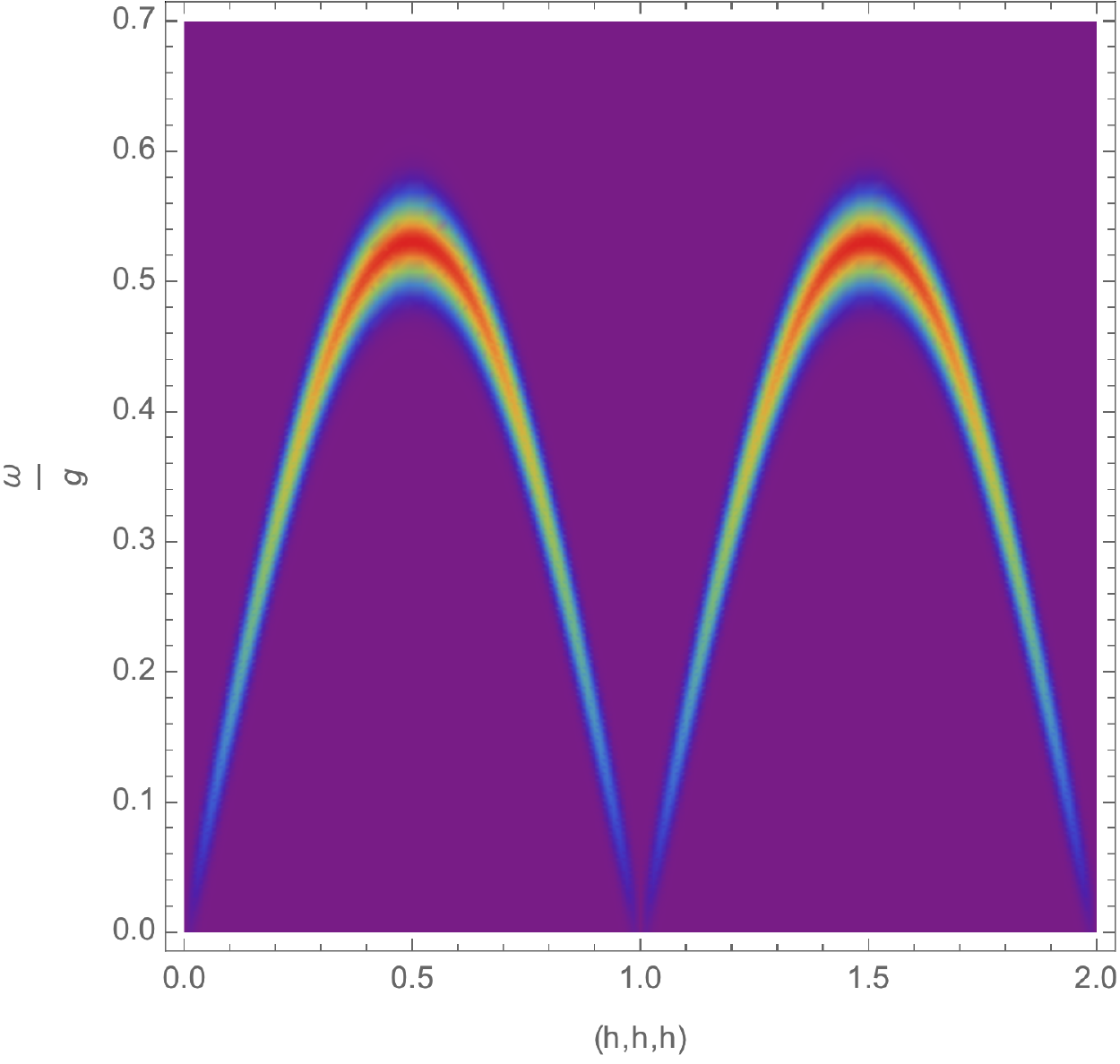}
\caption{
The unpolarised scattering intensity for $s=\frac{1}{2}$ at $T=0$ as might be seen in a typical neutron diffraction experiment along the direction $\mathbf{k}=\frac{2\pi}{a_0}(h,h,h)$. To simulate the finite resolution of the measuring apparatus, we have convoluted the intensity with a Gaussian of width $0.02g$.}
\label{fig: scattering at zero T}
\end{figure}

\begin{figure}
\centering
\includegraphics[width=0.9\columnwidth, angle=0]{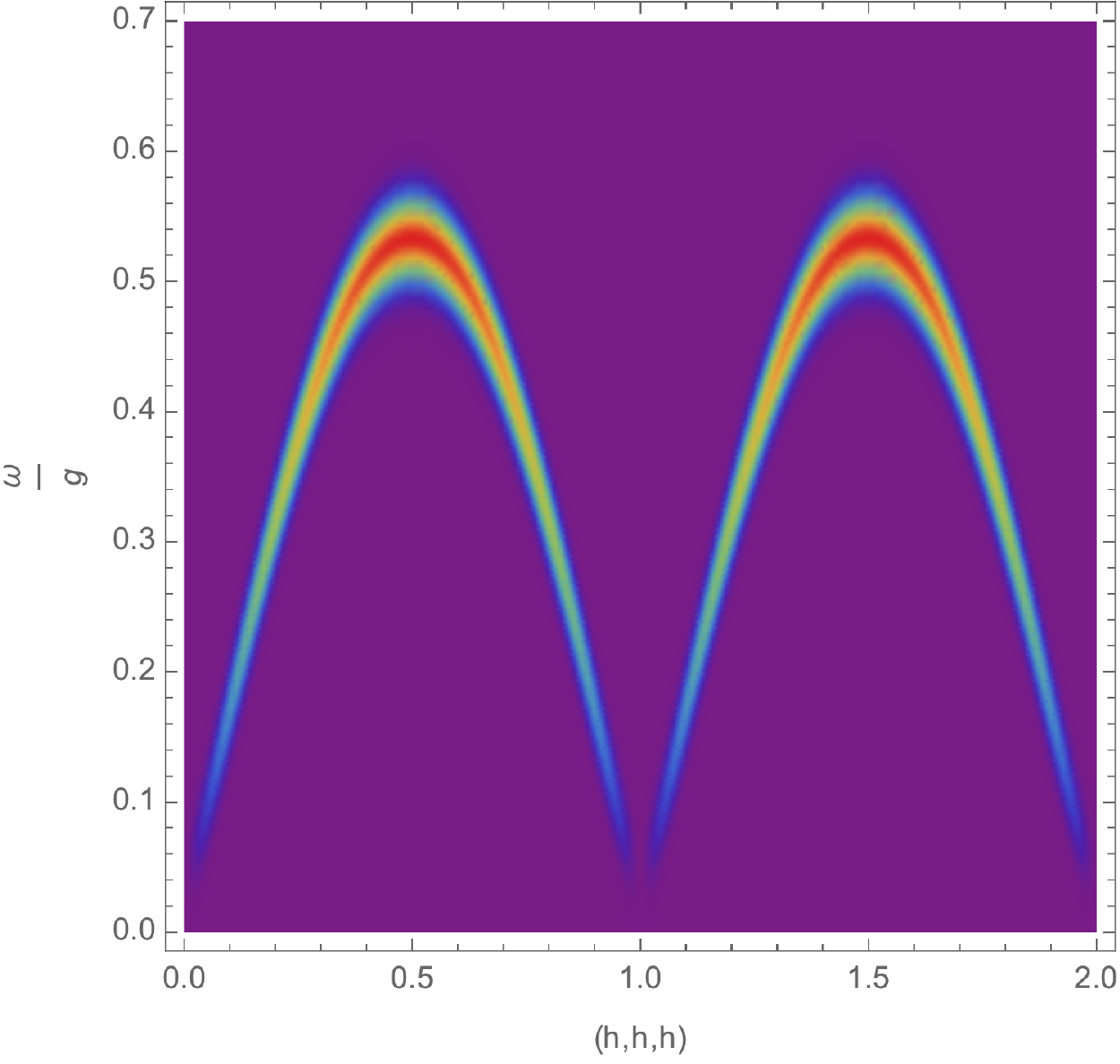}
\caption{
The unpolorised scattering intensity for $s=\frac{1}{2}$ at $T=0.03g$ as might be seen in a typical neutron diffraction experiment along the direction $\mathbf{k}=\frac{2\pi}{a_0}(h,h,h)$. Notice the development of a sizeable gap $\sim \omega_p$ at temperatures that are small by comparison with the photon bandwidth (and also the vison bandwidth evaluated for $s=\frac{1}{2}$). To simulate the finite resolution of the measuring apparatus, we have convoluted the intensity with a Gaussian of width $0.02g$.}
\label{fig: scattering at non-zero T}
\end{figure}
We should instead look for experimental signatures of the mass gap in the {\it unintegrated} neutron scattering spectra. The dynamical structure factor is defined as
\begin{eqnarray}
S^{\lambda\lambda}(\mathbf{k},\omega)=\int dt \; \langle \hat{S}^z_{\lambda}(\mathbf{k},t)\hat{S}^z_{\lambda}(-\mathbf{k},0)\rangle e^{i\omega t}.
\end{eqnarray}
We can analytically continue the imaginary time (Matsubara frequency) expectation values to obtain the dynamical structure factor
\begin{eqnarray}
&&S^{\lambda\lambda}(\mathbf{k},\omega)
=\frac{2}{1-e^{-\beta\omega}}\lim_{\epsilon\rightarrow 0}\chi(\mathbf{k}, \omega + i\epsilon)=
\nonumber\\
&&\frac{4g\pi\xi^2_\lambda(\mathbf{k})}{E_{\mathbf{k}}}\left(
n_B(E_{\mathbf{k}})\delta(\omega+E_{\mathbf{k}}) +\left(1+n_B(E_{\mathbf{k}})\right)\delta(\omega -E_{\mathbf{k}})
\right).
\nonumber\\
\end{eqnarray}
The total unpolarised scattering intensity is proportional to\cite{Shannon}
\begin{eqnarray}
I(\mathbf{k},\omega) = \sum_{\mu\nu} \left(
\hat{\mathbf{e}}_{\mu}\cdot\hat{\mathbf{e}}_{\nu}-
\frac{(\mathbf{k}\cdot\hat{\mathbf{e}}_{\mu})(\mathbf{k}\cdot\hat{\mathbf{e}}_{\nu})}{\mathbf{k}^2}
\right) S^{\mu\nu}(\mathbf{k},\omega),
\nonumber\\
\end{eqnarray}
where $S^{\mu\nu}(\mathbf{k},\omega)=\sum_{\lambda=1,2}S^{\lambda\lambda}(\mathbf{k},\omega)U^{\dagger}_{\mu\lambda}(\mathbf{k})U^{\dagger}_{\nu\lambda}(-\mathbf{k})$. The results are plotted in Fig.~\ref{fig: scattering at zero T} and Fig.~\ref{fig: scattering at non-zero T} and show a clear development of a mass gap at non-zero temperatures.

We compare our dynamical structure factor plots with the Quantum Monte Carlo calculations of Ref.~\onlinecite{Huang}. The reported bandwidth ($4.28g$) is significantly higher than the one we calculate ($0.6g$). However, we believe this discrepancy is mostly due to higher order corrections in the $1/s$ expansion. In fact, our previous work\cite{Our_Work} shows that the next order already renormalises the bandwidth from $0.6g$ to $1.6g$. Further, the plasma frequency and the speed of light should not be compared numerically in our calculation. This is because the calculation of the former 
is non-perturbative and includes all orders in $1/s$, whereas the speed of light has only been calculated to finite order in $1/s$. This does not impact the main observation though, which is that the photon acquires a mass gap equal to the plasma frequency. Our dynamical structure plots are in rough qualitative agreement with the work of Ref.~\onlinecite{Huang}. However, the Quantum Monte Carlo calculations do not have the required resolution to ascertain whether the photon dispersion is linear, let alone whether there is a small energy gap.

\subsection{Confinement of spinons}

We note that the {\it dynamical} mass gap generated by the vison plasma does not cause confinement of magnetic monopoles (spinons). Introducing a pair of magnetic monopoles into the system corresponds to choosing an appropriate {\it static} background field $A^0_{\lambda}(\mathbf{k})$, which only couples to the zero Matsubara frequency component of the dynamical gauge field $A_{\lambda}( \mathbf{k},\omega=0)$, which is not gapped. Hence, the vison plasma does not alter the Coulombic interaction between static magnetic monopoles introduced into the system. We refer the reader to appendix~\ref{App_Confinement} for a detailed mechanism of how a {\it static} mass gap, e.g. one generated by the condensation of visons in the ground state, could cause confinement of magnetic monopoles.

\section{Conclusion}
\label{Conclusion}

We have looked at how the gapped excitations of QSI known as visons can show up in its inelastic magnetic response. Because visons are sources of {\it emergent} electric field, they are not directly accessible to experimental probes. (Perhaps, magnetorestriction or Dzyaloshinskii-Moriya effects can couple real electric fields to the emergent electric field of the visons\cite{Attila_73, Attila_74, Attila_75}, in which case visons can have a direct experimental signature in electric reponse measurements.) In our theoretical analysis, we have looked at how visons impact the magnetic response, which can be directly probed in neutron scattering experiments.

To study how visons impact the magnetic response of QSI, we needed to capture their dynamics. We achieved this by looking at the ring-exchange Hamiltonian in the large-$s$ limit, but taking the perturbative analysis of our previous work\cite{Our_Work} further. We included non-perturbative corrections that correspond to the tunneling of visons between lattice sites. These quantum corrections also go beyond the recent classical description of Ref.~\onlinecite{Attila} and endow the visons with a finite mass (exponentially small in $s$), which we have calculated.

We have found that at low temperatures the visons form a diulte 3D Coulomb gas, which is always in the plasma phase, i.e. the visons are deconfined. We have investigated how this plasma interacts with the long-wavelength degrees of freedom of the gauge field. We have found that in the Debye limit, where a quadratic description is viable, plasma oscillations introduce a {\it dynamical} mass gap in the photon spectrum, which we have calculated as a function of temperature. 
This mass gap should be observable in inelastic magnetic response measurements. In particular, it should show up in energy-resolved neutron-scattering experiments and we have calculated the scattering intensity as would be seen in a typical experiment.
We have also compared our dynamical structure factor results with the recent quantum Monte Carlo (QMC) results of Ref.~\onlinecite{Huang}. These are in rough agreement, although, because of the limited resolution of QMC, the linear photon dispersion cannot be resolved, let alone a potential small gap in the spectrum.

We also show that the photon mass gap generated by the visons does not couple to the zero Matsubara frequency component of the gauge field, and hence does not result in confinement of {\it static} magnetic monopoles (spinons) introduced into the system.

Future questions to address, include the interactions between the vison plasma and {\it dynamical} magnetic monopoles introduced into the system and any potential experimental signatures of these interactions.

\acknowledgments{
This work was supported in part by EPSRC Grant Nos. EP/K028960/1 and EP/P013449/1, and by the 
EPSRC NetworkPlus on ``Emergence and Physics far from Equilibrium''. 
We gratefully acknowledge discussions with C. Castelnovo, A. Szab\'o, and G. Goldstein.
}

\appendix

\section{The longitudinal and transverse parts of the instanton solution in the continuum limit}
\label{app: instanton continuum}
It is very insightful to consider the continuum limit of the instanton solution in Eq.~\ref{eq: instanton solution}. We first write down the time-dependent part of the instanton solution as a sum of longitudinal and transverse parts
\begin{eqnarray}
E_{\mu}^{\rm inst.} (\mathbf{r}, \tau) - E_{\mu}^{\rm const.} (\mathbf{r})  = E_{\mu}^{\rm tran.} (\mathbf{r}, \tau)  +  E_{\mu}^{\rm long.} (\mathbf{r}, \tau) 
,
\nonumber\\
\end{eqnarray}
where by definition ${\rm div}_{\alpha} E^{\rm tran.} (\tau) = 0$ and ${\rm curl}_{ij} E^{\rm long.} (\tau)= 0$ everywhere. From the solution in Eq.~\ref{eq: instanton solution}, it follows that
\begin{eqnarray}
E_{\mu}^{\rm tran.} (\mathbf{r}, \tau) 
&=&
\frac{1}{N_s}
\sum_{\lambda=1,2, \mathbf{k} \in {\rm BZ}} 
\int_{-\infty}^{\infty} d\omega
\frac{-i \omega e^{-i\omega\tau +i \mathbf{k} \cdot \mathbf{r}} }{  \frac{2z\tilde{g}^2}{s^2} \xi_{\lambda}^2 (\mathbf{k}) + \frac{\omega^2}{2 }}
\nonumber\\
&&\times
 U^{\dagger}_{\mu \lambda} (\mathbf{k}) U_{\lambda\sigma}(\mathbf{k}) 
\nonumber\\
&\stackrel{|\mathbf{k}|a_0 \ll 1}{\approx}&
\int \frac{a_0^3 d^3 \mathbf{k}}{4\left(  2\pi \right)^3}
\int_0^{\infty} d\omega
\frac{-i \omega  e^{-i\omega\tau +i \mathbf{k} \cdot \mathbf{r}} }{ \left(  \frac{2z\tilde{g}^2 a_0^2}{s^2} |\mathbf{k}|^2 
+ \frac{\omega^2}{2} \right)  }
\nonumber\\
&&\times
\frac{3}{4} \left[  \hat{\mathbf{e}}_{\sigma} \cdot \hat{\mathbf{e}}_{\mu}  - 
\left(   \hat{\mathbf{k}} \cdot \hat{\mathbf{e}}_{\sigma}  \right) 
\left(      \hat{\mathbf{k}} \cdot \hat{\mathbf{e}}_{\mu}  \right)
 \right]
,
\nonumber\\
\end{eqnarray}
where we have used the identity
\begin{eqnarray}
&&\sum_{\lambda=1,2} U^{\dagger}_{\mu\lambda}(\mathbf{k}) U_{\lambda\sigma}(\mathbf{k})
\nonumber\\
&&=\frac{1}{\xi_1^2(\mathbf{k})}\sum_{\lambda} U^{\dagger}_{\mu\lambda}(\mathbf{k}) \xi^2_{\lambda}(\mathbf{k})U_{\lambda\sigma}(\mathbf{k})
\nonumber\\
&&=
\frac{1}{\xi_1^2(\mathbf{k})}\sum_{\nu}Z_{\mu\nu}(\mathbf{k})Z_{\nu\sigma}(\mathbf{k})
\nonumber\\
&&\stackrel{\mathbf{k}a_0\ll 1}{\approx}
\frac{-4}{|\mathbf{k}|^2a_0^2}
\sum_{\nu}\left(\mathbf{k}\cdot\Delta_{\mu\nu}\right)\left(\mathbf{k}\cdot\Delta_{\nu\sigma}\right)
\nonumber\\
&&=
\frac{3}{4} \left(
\left(\hat{\mathbf{e}}_{\mu}\cdot\hat{\mathbf{e}}_{\sigma}\right)-\left(\hat{\mathbf{k}}\cdot\hat{\mathbf{e}}_{\mu}\right)\left(\hat{\mathbf{k}}\cdot\hat{\mathbf{e}}_{\sigma}\right)
\right).
\nonumber\\
\end{eqnarray}
The transverse part of the instanton solution is transient in time with a power-law tail $\propto 1/|\tau|^3$ at long times. 

The longitudinal part of the electric field can also be extracted from the solution in Eq.~\ref{eq: instanton solution}
\begin{eqnarray}
E_{\mu}^{\rm long.} (\mathbf{r}, \tau) 
&=& 
-\frac{\pi}{N_s} {\rm sgn} (\tau) \sum_{\lambda=3,4, \mathbf{k} \in {\rm BZ}} e^{i \mathbf{k} \cdot \mathbf{r}}
\nonumber\\
 &&
\times U^{\dagger}_{\mu \lambda} (\mathbf{k}) U_{\lambda\sigma}(\mathbf{k}) 
\nonumber\\
&\stackrel{|\mathbf{k}|a_0 \ll 1}{\approx}&
\frac{3 \pi}{4} {\rm sgn} (\tau) \int \frac{a_0^3d^3 \mathbf{k}}{4\left(  2\pi \right)^3} e^{i \mathbf{k} \cdot \mathbf{r}}
\nonumber\\
&&
\times
\left(\frac{1}{3}+
\left(   \hat{\mathbf{k}} \cdot \hat{\mathbf{e}}_{\sigma}  \right) 
\left(      \hat{\mathbf{k}} \cdot \hat{\mathbf{e}}_{\mu}  \right)
\right)
\nonumber\\
&=&
\frac{3\pi a_0^3}{16}{\rm sgn} (\tau)\delta(\mathbf{r})+
A
\mathbf{E} (\mathbf{r},\tau) \cdot \hat{\mathbf{e}}_{\mu} 
,
\nonumber\\
\label{eq: B long.}
\end{eqnarray}
where the factor of $A=\frac{\sqrt{3}}{4} a_0^2$ links the lattice electric flux $E_{\mu}(\mathbf{r},\tau)$ with the continuum flux density $\mathbf{E}(\mathbf{r},\tau)$ and follows from approximating a sum over a large closed surface ${\rm S}$ by a surface integral $\sum_{(\mathbf{r},\mu) \in {\rm S}} \rightarrow \oint_{{\rm S}} \frac{dS}{A}$. $\mathbf{E} (\mathbf{r}, \tau)$ is the continuum electric field due to a dipole at $\mathbf{r}=\mathbf{0}$ with a dipole moment of $\pi{\rm sgn}(\tau)\mathbf{e}_{\sigma}$
\begin{eqnarray}
\mathbf{E} (\mathbf{r}, \tau) 
&=&
 -\pi i {\rm sgn}  (\tau) 
\mathbf{\nabla} \left[   \int \frac{d^3 \mathbf{k}}{\left(  2\pi \right)^3}
 \frac{\mathbf{k} \cdot \mathbf{e}_{\sigma}}{|\mathbf{k}|^2}  e^{i \mathbf{k} \cdot \mathbf{r}}   \right]  
\nonumber\\
&=&
\pi{\rm sgn}  (\tau) 
\mathbf{\nabla}
\left[\mathbf{e}_{\sigma}\cdot \mathbf{\nabla}\left( \frac{ 1}{|\mathbf{r}|} \right)
\right].
\end{eqnarray}
We have also used the identity
\begin{eqnarray}
\sum_{\lambda=3,4}U^{\dagger}_{\mu\lambda}(\mathbf{k}) U_{\lambda\sigma}(\mathbf{k})
&=&
\delta_{\mu\sigma}-\sum_{\lambda=1,2}U^{\dagger}_{\mu\lambda}(\mathbf{k}) U_{\lambda\sigma}(\mathbf{k})
\nonumber\\
&\stackrel{|\mathbf{k}|a_0\ll 1}{\approx}&\frac{3}{4}\left(\frac{1}{3}+
\left(   \hat{\mathbf{k}} \cdot \hat{\mathbf{e}}_{\sigma}  \right) 
\left(      \hat{\mathbf{k}} \cdot \hat{\mathbf{e}}_{\mu}  \right)
\right)
\nonumber\\
\end{eqnarray}
The longitudinal part of the electric field is uniquely determined by the charge distribution via Laplace's equation in Eq.~\ref{eq: Laplace}. Because $E_{\mu}^{\rm long.} (\mathbf{r}, \tau) $ has a lattice divergence of $-\pi{\rm sgn}(\tau)$ at $-\mathbf{e}_{\mu}/2$ and $\pi{\rm sgn} (\tau)$ at $\mathbf{e}_{\mu}/2$, it necessarily corresponds to the electric field due to a pair of charges $\pi$ and $-\pi$ at positions $-\mathbf{e}_{\mu}/2$ and $\mathbf{e}_{\mu}/2$ respectively for $\tau <0$ that switch positions for $\tau >0$. We therefore obtain a dipole field as above in the continuum limit. It also follows that, to describe the hopping of a vison of charge $2\pi$ at $\tau=0$, the constant background electric field $E_{\mu}^{\rm const.} (\mathbf{r}, \tau)$ has to be the longitudinal field due to a pair of charges $\pi$ at positions $\pm \mathbf{e}_{\mu}/2$.

\section{The smooth instanton solution
\label{app: smooth instanton}}

We begin with the instanton solution in Eq.~\ref{eq: instanton solution} (in the following the instanton solution is given relative to the constant and divergenceful background field $E_{\lambda}^{\rm const.} (\mathbf{k})$)

\begin{eqnarray}
E^{\rm inst.}_{\lambda}(\mathbf{k},\tau)= 
\frac{d}{d \tau}
\left[
\frac{1}{\sqrt{N_s} \beta}
\sum_{\omega}
\frac{\pi U_{\lambda \sigma}(\mathbf{k})}{\frac{2z\tilde{g}^2}{s^2}\xi_{\lambda}^2(\mathbf{k})
+\frac{\omega^2}{2}} e^{-i \omega \tau}
\right].
\nonumber\\
\end{eqnarray}
Taking the low temperature limit $\beta\tilde{g}\sqrt{z}/s \rightarrow \infty$, we obtain
\begin{eqnarray}
E^{\rm inst.}_{\lambda}(\mathbf{k},\tau)
&=& 
\frac{d}{d \tau}
\int_{-\infty}^{\infty}
\frac{d \omega}{2 \pi \sqrt{N_s}}
\frac{\pi U_{\lambda \sigma}(\mathbf{k})}{\frac{2z\tilde{g}^2}{s^2}\xi_{\lambda}^2(\mathbf{k})
+\frac{\omega^2}{2}} e^{-i \omega \tau}
\nonumber\\
&=&
\frac{s \pi U_{\lambda \sigma} (\mathbf{k})}{2\tilde{g} \sqrt{N_sz} \xi_{\lambda} (\mathbf{k})} 
\frac{d}{d \tau} \left[
e^{-|\tau| \frac{2\tilde{g}\sqrt{z}|\xi_{\lambda}(\mathbf{k})|}{s}} \right]
\nonumber\\
&=&
-\frac{\pi {\rm sgn} (\tau) U_{\lambda \sigma} (\mathbf{k})}{\sqrt{N_s}} e^{-|\tau| \frac{2\tilde{g}\sqrt{z}|\xi_{\lambda}(\mathbf{k})|}{s}} .
\end{eqnarray}
The smooth instanton solution in Eq.~\ref{eq: smooth instanton} is given by 
\begin{eqnarray}
&&E^{\rm sm.} _{\lambda}(\mathbf{k}, \tau) 
=
\nonumber\\
&&
\frac{\pi U_{\lambda \sigma} (\mathbf{k})}{\sqrt{N_s}} \left[
-{\rm sgn}(\tau) e^{-|\tau| \frac{2\tilde{g}\sqrt{z}|\xi_{\lambda}(\mathbf{k})|}{s}} + {\rm sgn}(\tau) + 1
\right],
\nonumber\\
\end{eqnarray}
and we can see that the jump in the field has been removed.
Straightforward differentiation gives
\begin{eqnarray}
\ddot{E}^{\rm sm.} _{\lambda}(\mathbf{k}, \tau) 
&=&
-\frac{\pi {\rm sgn} (\tau) U_{\lambda \sigma} (\mathbf{k})}{\sqrt{N_s}} 
\frac{4z\tilde{g}^2 \xi^2_{\lambda} (\mathbf{k})}{s^2}
 e^{-|\tau| \frac{2\tilde{g}\sqrt{z}|\xi_{\lambda}(\mathbf{k})|}{s}} .
\nonumber\\
\end{eqnarray}
The smooth instanton solution is the stationary solution of the action in Eq.~\ref{eq: instanton action}. Minimising that action with respect to variations in the dynamical, transverse part of the field $E_{\lambda=1,2} (-\mathbf{k}, \tau)$ we obtain the following saddle point equations
\begin{eqnarray}
 \frac{-2s^2}{4z\tilde{g} \xi^2_{\lambda} (\mathbf{k})} \ddot{E}_{\lambda} (\mathbf{k}, \tau) + \tilde{g}
\sum_{\mathbf{r}, \mu} V' \left[  E_{\mu} (\mathbf{r}, \tau)   \right] \frac{\delta E_{\mu} (\mathbf{r}, \tau)}{\delta E_{\lambda} (-\mathbf{k}, \tau)} =0, 
\label{eq: smooth saddle point}
\nonumber\\
\end{eqnarray}
for all $\mathbf{k}$ and $\lambda=1,2$, where
\begin{eqnarray}
\frac{\delta E_{\mu} (\mathbf{r}, \tau)}{\delta E_{\lambda} (-\mathbf{k}, \tau)} =
\frac{1}{\sqrt{N_s}} U^{\dagger}_{\mu \lambda} (-\mathbf{k}) e^{-i \mathbf{k}\cdot \mathbf{r}}
\end{eqnarray}
and
\begin{eqnarray}
V' \left[  E_{\mu} (\mathbf{r}, \tau) \right] = 2 E_{\mu} (\mathbf{r}, \tau) 
- 2\pi \delta_{\mu\sigma} \delta_{\mathbf{r},\mathbf{0}} \left[ {\rm sgn}(\tau) + 1 \right].
\nonumber\\
\end{eqnarray}
The first derivative of the continued parabolic potential $V' \left[  E_{\mu} (\mathbf{r}, \tau) \right]$ is given by $2 E_{\mu} (\mathbf{r}, \tau)$, when $ |E_{\mu} (\mathbf{r}, \tau)| \leq \pi$. This is true in the case of the above smooth instanton solution $E^{\rm sm.} _{\mu}(\mathbf{r}, \tau)$, everywhere except for the single electric field $E_{\mu=\sigma} (\mathbf{r} = \mathbf{0}, \tau)$, which lies between $\pi$ and $3\pi$ for positive imaginary times. In this case $V' \left[  E_{\mu} (\mathbf{r}, \tau) \right]= 2 E_{\mu} (\mathbf{r}, \tau) - 4\pi$, thus justifying the above expression.
It follows that the smooth instanton solution $E^{\rm sm.} _{\mu}(\mathbf{r}, \tau)$ satisfies the saddle point equation in Eq.~\ref{eq: smooth saddle point} 
\begin{eqnarray}
0 
&=&
-2\ddot{E}^{\rm sm.}_{\lambda} (\mathbf{k},\tau) \frac{s^2}{4z\tilde{g}\xi_{\lambda}^2(\mathbf{k})} + 2\tilde{g}E^{\rm sm.}_{\lambda} (\mathbf{k}, \tau) 
\nonumber\\
&&- \frac{2\pi\tilde{g}}{\sqrt{N_s}} U_{\lambda \sigma}  (\mathbf{k}) \left[  {\rm sgn}(\tau) + 1 \right].
\end{eqnarray}

We now turn to proving that $\dot{E}^{\rm sm.}_{\lambda} (\mathbf{k}, \tau)$ is the zero-energy mode of the action in Eq.~\ref{eq: instanton fluctuations} describing fluctuations around the stationary smooth instanton solution 
\begin{eqnarray}
&&\delta \mathcal{S}_{\rm tran} \left[   \delta E_{\lambda} (\mathbf{k}, \tau)  \right] 
=
\int_{-\beta/2}^{\beta/2} d\tau
\sum_{\mathbf{k} \in {\rm BZ}, \lambda=1,2} 
 \frac{ s^2  |\delta \dot{E}_{\lambda} (\mathbf{k}, \tau)|^2  }{4z\tilde{g}\xi_{\lambda}^2(\mathbf{k})} 
\nonumber\\
&&+ \int_{-\beta/2}^{\beta/2} d\tau\;
 \frac{\tilde{g}}{2}\sum_{\alpha\beta} V'' \left(E^{\rm sm.}_{\alpha\beta}(\tau)\right) \delta E^2_{\alpha\beta} (\tau).
\nonumber\\
\end{eqnarray}
The zero-mode of the above action satisfies the following differential equation for $\lambda=1,2$:
\begin{eqnarray}
0
&=&
-\frac{s^2}{4z\tilde{g} \xi_{\lambda}^2(\mathbf{k})} \delta \ddot{E}_{\lambda} (\mathbf{k},\tau) 
\nonumber\\
&&+ \frac{\delta E_{\mu} (\mathbf{r}, \tau)}{\delta E_{\lambda} (-\mathbf{k}, \tau)}\frac{\tilde{g}}{2}\sum_{\alpha\beta}  V'' \left[  E^{\rm sm.}_{\alpha\beta} (\tau)  \right] \delta E_{\alpha\beta} (\tau)
\nonumber\\
\end{eqnarray}
We see that $\delta E_{\alpha\beta} (\tau) = \dot{E}^{\rm sm.}_{\alpha\beta} (\tau)$ is a solution of the equation, because
\begin{eqnarray}
&&-\frac{s^2}{4z\tilde{g} \xi_{\lambda}^2(\mathbf{k})} \dddot{E}^{\rm sm.}_{\lambda} (\mathbf{k}) 
+
\nonumber\\
&&\frac{\delta E_{\mu} (\mathbf{r}, \tau)}{\delta E_{\lambda} (-\mathbf{k}, \tau)}\frac{\tilde{g}}{2}\sum_{\alpha\beta}  V'' \left[  E^{\rm sm.}_{\alpha\beta} (\tau)  \right] \dot{E}^{\rm sm.}_{\alpha\beta} (\tau)
\end{eqnarray}
is proportional to the time derivative of the left-hand side of Eq.~\ref{eq: smooth saddle point}.

\section{Instanton measure
\label{app: instanton measure}}

The fluctuation action in Eq.~\ref{eq: instanton fluctuations} can be written as follows
\begin{eqnarray}
\delta S &=& \sum_{\omega}\sum_{\mathbf{k},\lambda=1,2}
\left(\tilde{g}+\frac{s^2\omega^2}{4z\tilde{g}\xi_{\lambda}^2(\mathbf{k})}
\right)|\delta E_{\lambda}(\mathbf{k},\omega)|^2
\nonumber\\
&&-\frac{2\pi \tilde{g}}{\dot{E}^{\rm sm.}_{\sigma}(\mathbf{0},0)}
\delta E_{\sigma}^2(\mathbf{0},0),
\end{eqnarray}
where
\begin{eqnarray}
\dot{E}^{\rm sm.}_{\sigma}(0,\mathbf{0}) &=& 
\frac{2\sqrt{z}\pi \tilde{g}}{sN_s} \sum_{\mathbf{k},\lambda}
U^{\dagger}_{\sigma\lambda}(\mathbf{k})|\xi_{\lambda}(\mathbf{k})| U_{\lambda\sigma}(\mathbf{k})
\nonumber\\
&=&
\frac{2\sqrt{z}\pi \tilde{g}}{sN_s}\sum_{\mathbf{k} \in {\rm B.Z.},\mu}
\frac{Z_{\sigma \mu}(\mathbf{k}) Z_{\mu \sigma}(\mathbf{k}) }{|\xi_1(\mathbf{k})|}
\nonumber\\
&\equiv&\frac{C_3^2\tilde{g}}{s},
\label{eq: field velocity}
\end{eqnarray}
and $C_3=5.13$. It is convenient to express the fluctuation action in matrix form
\begin{eqnarray}
\delta S = \mathbf{E}^{\dagger}   \tilde{\mathbf{K}} \mathbf{E}=
\mathbf{E}^{\dagger} \left(  \mathbf{K} -\lambda \mathbf{v}\mathbf{v}^{\dagger} \right)\mathbf{E},
\end{eqnarray}
where
\begin{eqnarray}
\left[\mathbf{E}\right]_{\lambda\omega\mathbf{k}}&=& \delta E_{\lambda}(\omega,\mathbf{k}),
\nonumber\\
\left[\mathbf{v}\right]_{\lambda\omega\mathbf{k}} &=&\frac{1}{\sqrt{N_s\beta}} U_{\lambda\sigma}(\mathbf{k}),
\nonumber\\
\left[\mathbf{K}\right]_{\lambda\omega\mathbf{k},\lambda'\omega'\mathbf{k}'}
&=& \delta_{\lambda\omega\mathbf{k},\lambda'\omega'\mathbf{k}'}\left(\tilde{g}+\frac{s^2\omega^2}{4z\tilde{g}\xi_{\lambda}^2(\mathbf{k})}\right),
\nonumber\\
\lambda &=& \frac{2\pi s}{C_3^2}=\frac{1}{\mathbf{v}^{\dagger}\mathbf{K}^{-1}\mathbf{v}},
\label{eq: matrix mapping}
\end{eqnarray}
where the final identity follows from Eq.~\ref{eq: field velocity}. 

The instanton measure can be obtained by working out the contribution of the one-instanton sector to the partition function relative to the zero-instanton sector
\begin{eqnarray}
&&\frac{\int\prod_i d\tilde{\xi}_i \sqrt{\langle \tilde{\Psi}_i |\tilde{\Psi}_i\rangle} e^{-\frac{1}{2} \sum_{i\neq 0}\tilde{\lambda}_i \tilde{\xi}_i^2 \langle \tilde{\Psi}_i |\tilde{\Psi}_i\rangle}}
{\int\prod_i d\xi_i \sqrt{\langle \Psi_i |\Psi_i\rangle}
e^{-\frac{1}{2} \sum_{i}\lambda_i \xi_i^2 \langle \Psi_i |\Psi_i\rangle}}
\nonumber\\
&&=
\int d\tilde{\xi}_0 \sqrt{\frac{\langle \tilde{\Psi}_0| \tilde{\Psi}_0 \rangle}{2\pi}  }
\sqrt{\frac{{\rm det}\; \mathbf{K}}{{\rm det}' \tilde{\mathbf{K}}}},
\label{eq: instanton measure formula}
\end{eqnarray}
where we have written down $\delta E( \mathbf{r},\tau)=\sum_i \xi_i \Psi_i( \mathbf{r},\tau)=\sum_i \tilde{\xi}_i \tilde{\Psi}_i( \mathbf{r},\tau)$ in terms of the {\it real} eigenvectors of $\mathbf{K}$ and $\tilde{\mathbf{K}}$ respectively in the $( \mathbf{r},\tau)$ basis and the zero eigenvalue is excluded from the determinant of $\tilde{\mathbf{K}}$.
From
\begin{eqnarray}
\tilde{\mathbf{K}}= \mathbf{K}\left(\mathbf{1} - \lambda\mathbf{K}^{-1}\mathbf{v}\mathbf{v}^{\dagger}
\right),
\end{eqnarray}
it follows that
\begin{eqnarray}
\frac{{\rm det}\left(\tilde{\mathbf{K}}\right)}{ {\rm det}\left(\mathbf{K}\right)}=
{\rm det}\left(\mathbf{1} - \lambda\mathbf{K}^{-1}\mathbf{v}\mathbf{v}^{\dagger}
\right).
\end{eqnarray}
The matrix $\left(\mathbf{1} + \lambda\mathbf{K}^{-1}\mathbf{v}\mathbf{v}^{\dagger}
\right)$ has $N-1$ eigenvectors perpendicular to $\mathbf{v}$ with eigenvalue $1$ and an eigenvector $\mathbf{K}^{-1}\mathbf{v}$ with eigenvalue equal to $\left(1-\lambda\mathbf{v}^{\dagger}\mathbf{K}^{-1}\mathbf{v}\right)$. Therefore
\begin{eqnarray}
\frac{{\rm det}\left(\tilde{\mathbf{K}}\right)}{ {\rm det}\left(\mathbf{K}\right)}=\left(1-\lambda\mathbf{v}^{\dagger}\mathbf{K}^{-1}\mathbf{v}\right).
\end{eqnarray}
When $\lambda$ takes on its physical value given by Eq.~\ref{eq: matrix mapping} and equal to $\left(\mathbf{v}^{\dagger}\mathbf{K}^{-1}\mathbf{v}\right)^{-1}$, ${\rm det}\; \tilde{\mathbf{K}}$ and the above ratio vanish. To avoid this, we will perturb $\lambda$ from its physical value as follows
\begin{eqnarray}
\lambda=\frac{1}{\mathbf{v}^{\dagger}\mathbf{K}^{-1}\mathbf{v}} - \delta\lambda.
\end{eqnarray}
The matrix $\tilde{\mathbf{K}}$ changes as a result by $\delta\tilde{\mathbf{K}}=\delta\lambda\mathbf{v}\mathbf{v}^{\dagger}$, and
standard perturbation theory gives the shift of the zero eigenvalue of $\tilde{\mathbf{K}}$ 
\begin{eqnarray}
\delta\tilde{\lambda}_0=\frac{\langle\mathbf{K}^{-1}\mathbf{v} |\delta\tilde{\mathbf{K}}|\mathbf{K}^{-1}\mathbf{v}\rangle}{\langle\mathbf{K}^{-1}\mathbf{v}|\mathbf{K}^{-1}\mathbf{v}\rangle}
=
\frac{\left(\mathbf{v}^{\dagger}\mathbf{K}^{-1}\mathbf{v}\right)^2\delta\lambda}{\mathbf{v}^{\dagger}\mathbf{K}^{-2}\mathbf{v}}.
\end{eqnarray}
With the perturbed $\lambda$, we can now evaluate
\begin{eqnarray}
\frac{{\rm det}'\left(\tilde{\mathbf{K}}\right)}{ {\rm det}\left(\mathbf{K}\right)}=\frac{{\rm det}\left(\tilde{\mathbf{K}}\right)}{\delta\tilde{\lambda}_0{\rm det}\left(\mathbf{K}\right)}=\frac{\mathbf{v}^{\dagger}\mathbf{K}^{-1}\mathbf{v}\delta\lambda}{\delta\tilde{\lambda}_0}=\frac{\mathbf{v}^{\dagger}\mathbf{K}^{-2}\mathbf{v}}{\mathbf{v}^{\dagger}\mathbf{K}^{-1}\mathbf{v}},
\nonumber\\
\end{eqnarray}
and then take the limit $\delta\lambda\rightarrow 0$, where
\begin{eqnarray}
\mathbf{v}^{\dagger}\mathbf{K}^{-2}\mathbf{v}&=&\frac{1}{N_s\beta}\sum_{\omega}\sum_{\mathbf{k},\lambda} U^{\dagger}_{\sigma\lambda}(\mathbf{k}) \left(\frac{1}{\tilde{g}+\frac{\omega^2s^2}{4z\tilde{g}\xi^2_{\lambda}(\mathbf{k})}}\right)^2U_{\lambda\sigma}(\mathbf{k})
\nonumber\\
&=&
\frac{C^2_3}{4\pi \tilde{g}s},
\end{eqnarray}
and we have summed over $\omega$ in the $\beta\tilde{g}\sqrt{z}/s\rightarrow\infty$ limit and used the identity quoted in Eq.~\ref{eq: field velocity}.

We also calculate the norm of the zero mode
\begin{eqnarray}
\langle \tilde{\Psi}_0|\tilde{\Psi}_0\rangle&=&\sum_{\mathbf{k},\lambda}\int_{-\infty}^{\infty} |\dot{E}^{\rm sm.}_{\lambda}(\tau,\mathbf{k})|^2 d\tau
\nonumber\\
&=&\frac{2\sqrt{z}\tilde{g}\pi^2}{N_s s}\sum_{\mathbf{k},\lambda}
\frac{U^{\dagger}_{\sigma\lambda}\xi^2_\lambda(\mathbf{k})U_{\lambda\sigma}(\mathbf{k})}{|\xi_1(\mathbf{k})|}
\nonumber\\
&=&\frac{C_3^2\pi \tilde{g}}{s},
\end{eqnarray}
where we have again used the identity quoted in Eq.~\ref{eq: field velocity}.
Inserting everything into Eq.~\ref{eq: instanton measure formula}, the instanton measure becomes
\begin{eqnarray}
\int \frac{C_3\tilde{g}\:d\tilde{\xi_0}}{\sqrt{s}},
\end{eqnarray}
where $\xi_0$ specifies the position of the instanton in imaginary time.

\section{Confinement of spinons}
\label{App_Confinement}
In this section, we demonstrate how a non-dynamical mass gap, i.e. one that couples to the zero Matsubara frequency component, would generate confinement.

We introduce two oppositely charged monopoles situated on 'up' tetrahedra with centres $\mathbf{r} \pm N\left( \mathbf{e}_1-\mathbf{e}_2\right)$. This results in a line of background flux linking them
\begin{eqnarray}
&&\sqrt{N_s}B^{0}_{\lambda} (\mathbf{k}) = 
\sum_{\mathbf{r},\mu}e^{-i \mathbf{k} \cdot (\mathbf{r} + \mathbf{e}_{\mu}/2)}U_{\lambda \mu}(\mathbf{k}) B_{\mu}(\mathbf{r} + \mathbf{e}_{\mu}/2) \nonumber\\
&&= \sum_{n=-N}^{N}
\left(U_{\lambda 1} (\mathbf{k}) e^{-i\mathbf{k}\cdot(n\mathbf{a}+\mathbf{e}_1/2)}
-U_{\lambda 2}(\mathbf{k})e^{-i\mathbf{k}\cdot(n\mathbf{a}+\mathbf{e}_2/2)}
\right)
\nonumber\\
&&=
\left(U_{\lambda 1}(\mathbf{k}) e^{-i\mathbf{k}\cdot \mathbf{e}_1/2}
-U_{\lambda 2}(\mathbf{k}) e^{-i\mathbf{k}\cdot \mathbf{e}_2/2}
\right)
\frac{\sin \mathbf{k}\cdot\mathbf{a}(N+\frac{1}{2})}
{\sin \mathbf{k}\cdot\mathbf{a}},
\nonumber\\
\end{eqnarray}
where $\mathbf{a}=\mathbf{e}_1-\mathbf{e}_2$. By Eq.~\ref{eq: background magnetic field}, the non-integer background vector potential is then given by $A^0_{\lambda} (\mathbf{k})= B^{0}_{\lambda}(\mathbf{k})/\xi_{\lambda}(\mathbf{k})$ for $\lambda=1,2$. In its presence, the static part of the action in Eq.~\ref{eq: gapped gauge action} becomes
\begin{eqnarray}
S/\beta = \sum_{\mathbf{k}, \lambda=1,2} \frac{\tilde{g}za_0^2}{s^2} \left(\mathbf{k}^2
|A_{\lambda}(\mathbf{k})+A^0_{\lambda}(\mathbf{k})|^2
+\xi^{-2} |A_{\lambda}(\mathbf{k})|^2
\right),
\nonumber\\
\end{eqnarray}
where $\xi^{-2}$ is a static mass gap that could be generated by vison condensation in the ground state. Integrating out the field $A_{\lambda}(\mathbf{k})$, we obtain
\begin{eqnarray}
S/\beta= \frac{\tilde{g} za_0^2}{s^2} \sum_{\mathbf{k}, \lambda} \frac{\mathbf{k}^2\xi^{-2}}{\mathbf{k}^2+\xi^{-2}} |A^0_{\lambda} (\mathbf{k})|^2.
\end{eqnarray}
We can see that in the absence of a static mass gap, i.e. in the limit $\xi\rightarrow\infty$, there is no energy cost arising from the transverse part of the background field $B^0_{\lambda}(\mathbf{k})$. There is only a Coulombic energy cost arising from its longitudinal part, i.e. $\vartriangle_{ij}\psi$ in Eq.~\ref{eq: background magnetic field}. Substituting in for $A^0_{\lambda}(\mathbf{k})$ and integrating along $\mathbf{k}\cdot \hat{\mathbf{a}}$ first, we obtain
\begin{eqnarray}
S/\beta &=& \frac{\tilde{g}za_0^3\pi n}{|\mathbf{a}|} \int_{|\mathbf{k}|< \Lambda} \frac{d^3 \mathbf{k}}{(2\pi)^3}
\frac{\xi^{-2}}{\mathbf{k}^2+\xi^{-2}}\frac{|\mathbf{a}|\sin^2 n\mathbf{k}\cdot\mathbf{a}}{\pi n(\mathbf{k}\cdot\mathbf{a})^2}
\nonumber\\
&&\times
|U_{\lambda 1}(\mathbf{k})e^{-i\mathbf{k}\cdot{\mathbf{e}_1/2}}-U_{\lambda 2}(\mathbf{k})e^{-i\mathbf{k}\cdot{\mathbf{e}_2/2}}|^2
\nonumber\\
&=&\frac{2\pi \tilde{g}z a_0^3 n}{|\mathbf{a}|} \int_{|\mathbf{k}|<\Lambda} \frac{d^2 \mathbf{k}}{(2\pi)^3} 
\frac{\xi^{-2}}{\mathbf{k}^2+\xi^{-2}}
\nonumber\\
&\approx& n \frac{\tilde{g}z(a_0/\xi)^2\ln\Lambda\xi}{\sqrt{2}\pi},
\end{eqnarray}
where we have worked in the limit $\Lambda a_0 \ll 1$ and used the fact that $\lim_{n \rightarrow \infty}\left(\frac{|\mathbf{a}|}{\pi n} \frac{\sin^2n\mathbf{k}\cdot\mathbf{a}}{(\mathbf{k}\cdot\mathbf{a})^2}\right)=\delta(\mathbf{k}\cdot\hat{\mathbf{a}})$ and $|U_{\lambda 1}(\mathbf{k})e^{-i\mathbf{k}\cdot{\mathbf{e}_1/2}}-U_{\lambda 2}(\mathbf{k})e^{-i\mathbf{k}\cdot{\mathbf{e}_2/2}}|^2=2$ at $\mathbf{k}\cdot\hat{\mathbf{a}}=0$. The above result shows that a static mass gap generates a constant confining force equal to \begin{eqnarray}
F=\frac{\tilde{g}z(a_0/\xi)^2\ln\Lambda\xi}{\sqrt{2}\pi a_0}.
\end{eqnarray}

\section{Plasma oscillations in the continuum limit
\label{app: continuum}}

We begin with the action in Eq.~\eqref{partition and action} and take the continuum limit as follows
\begin{eqnarray}
&&A_{\alpha\beta}(\tau)\rightarrow \mathbf{A}(\mathbf{r},\tau)\cdot\left( \mathbf{r}_{\beta}-\mathbf{r}_{\alpha}  \right),
\nonumber\\
&&E_{\alpha\beta}(\tau)\rightarrow \mathbf{E}(\mathbf{r},\tau)\cdot\left( \mathbf{r}_{\beta}-\mathbf{r}_{\alpha}  \right),
\nonumber\\
&&\mathcal{S}=
\int_0^{\beta}d\tau \int \frac{d^3 \mathbf{r} }{a_0} \left(
i\mathbf{E} \cdot \dot{\mathbf{A}} 
+ \tilde{g} \mathbf{E}^2 + \frac{\tilde{g}za_0^2}{s^2} (\nabla \times \mathbf{A})^2
\right).
\nonumber\\
\end{eqnarray}
We can also consider adding a vison current term to the action which couples to the magnetic vector potential
\begin{eqnarray}
\delta \mathcal{S} =-\int d^3 \mathbf{r}
\;
\mathbf{J}\cdot \mathbf{A}.
\end{eqnarray}
Moving from imaginary to real time (which we measure in units of $\hbar$), we obtain the saddle point equations of the action, i.e. Maxwell's equations
\begin{eqnarray}
\dot{\mathbf{A}} &=& -2\tilde{g} \mathbf{E},
\nonumber\\
\dot{\mathbf{E}} &=& \frac{2\tilde{g}a_0^2z}{s^2} \nabla \times \left(\nabla \times \mathbf{A}\right) - \mathbf{J}a_0.
\label{eq: continuum equations of motion}
\end{eqnarray}
In the Debye limit $e^{\beta \mu_V} \gg 1$, when there are many visons inside the Debye volume, the visons will respond to the gauge field coherently and we can use a hydrodynamic description. The charges accelerate in the presence of the electric field according to
\begin{eqnarray}
\dot{\mathbf{J}}=-2(2\pi\hbar)^2n\dot{\mathbf{A}}/m^{\ast},
\end{eqnarray}
where $n=8e^{-\beta\mu_V}a_0^{-3}$ is the vison number density and $(2\pi\hbar)$ is the vison charge.
Differentiating the second equation in Eq.~\ref{eq: continuum equations of motion} with respect to time, substituting in for $\dot{\mathbf{A}}$ with the help of the first equation and using $\dot{\mathbf{J}}=-2(2\pi\hbar)^2n\dot{\mathbf{A}}/m^{\ast}$, we obtain the plasma equation
\begin{eqnarray}
c^2 \nabla^2 \mathbf{E} = \ddot{\mathbf{E}} + \omega_p^2\mathbf{E},
\end{eqnarray}
where $c^2=\frac{4\tilde{g}^2a_0^2z}{s^2}$, $\omega_p^2=\frac{128\pi^2\hbar^2e^{-\beta\mu_V}\tilde{g}}{m^{\ast}a_0^2}$, and we have assumed plasma neutrality so that $\nabla\cdot\mathbf{E} =-\nabla^2\varphi =0$ everywhere. This result agrees with the derivation presented in the main body of this paper.


\begin{thebibliography}{1}%
\makeatletter
\providecommand \@ifxundefined [1]{%
 \@ifx{#1\undefined}
}%
\providecommand \@ifnum [1]{%
 \ifnum #1\expandafter \@firstoftwo
 \else \expandafter \@secondoftwo
 \fi
}%
\providecommand \@ifx [1]{%
 \ifx #1\expandafter \@firstoftwo
 \else \expandafter \@secondoftwo
 \fi
}%
\providecommand \natexlab [1]{#1}%
\providecommand \enquote  [1]{``#1''}%
\providecommand \bibnamefont  [1]{#1}%
\providecommand \bibfnamefont [1]{#1}%
\providecommand \citenamefont [1]{#1}%
\providecommand \href@noop [0]{\@secondoftwo}%
\providecommand \href [0]{\begingroup \@sanitize@url \@href}%
\providecommand \@href[1]{\@@startlink{#1}\@@href}%
\providecommand \@@href[1]{\endgroup#1\@@endlink}%
\providecommand \@sanitize@url [0]{\catcode `\\12\catcode `\$12\catcode
  `\&12\catcode `\#12\catcode `\^12\catcode `\_12\catcode `\%12\relax}%
\providecommand \@@startlink[1]{}%
\providecommand \@@endlink[0]{}%
\providecommand \url  [0]{\begingroup\@sanitize@url \@url }%
\providecommand \@url [1]{\endgroup\@href {#1}{\urlprefix }}%
\providecommand \urlprefix  [0]{URL }%
\providecommand \Eprint [0]{\href }%
\providecommand \doibase [0]{http://dx.doi.org/}%
\providecommand \selectlanguage [0]{\@gobble}%
\providecommand \bibinfo  [0]{\@secondoftwo}%
\providecommand \bibfield  [0]{\@secondoftwo}%
\providecommand \translation [1]{[#1]}%
\providecommand \BibitemOpen [0]{}%
\providecommand \bibitemStop [0]{}%
\providecommand \bibitemNoStop [0]{.\EOS\space}%
\providecommand \EOS [0]{\spacefactor3000\relax}%
\providecommand \BibitemShut  [1]{\csname bibitem#1\endcsname}%
\let\auto@bib@innerbib\@empty
\bibitem [{Note1()}]{Note1}%
  \BibitemOpen
  \bibinfo {note} {In this limit, the typical instanton separation along $\tau
  $ will be small by comparison with $\beta $, so that finite-size effects
  associated with finite temperature are negligible.}\BibitemShut {Stop}%
\end{thebibliography}%


\begin{thebibliography}{10}

\bibitem{Gingras} M. J. P. Gingras and P. A. McClarty, Rep. Prog. Phys. {\bf 77}, 056501 (2014).
\bibitem{Knolle}
J. Knolle and R Moessner, Annu. Rev. Condens. Matter Phys. {\bf 10}, 451, (2019).
\bibitem{Attila_4}
A. Kitaev, Ann. Phys. {\bf 321}, 2 (2006).
\bibitem{Attila_5}
A. Banerjee, C. A. Bridges, J.-Q. Yan, A. A. Aczel, L. Li, M. B.
Stone, G. E. Granroth, M. D. Lumsden, Y. Yiu, J. Knolle, S.
Bhattacharjee, D. Kovrizhin, R. Moessner, D. A. Tennant, D. G.
Mandrus, and S. E. Nagler, Nat. Mater. {\bf 15}, 733 (2016).
\bibitem{Attila_6}
A. Banerjee, J. Yan, J. Knolle, C. A. Bridges, M. B. Stone,
M. D. Lumsden, D. G. Mandrus, D. A. Tennant, R. Moessner,
and S. E. Nagler, Science {\bf 356}, 1055 (2017).
\bibitem{Attila_7}
S.-H. Do, S.-Y. Park, J. Yoshitake, J. Nasu, Y. Motome, Y. S.
Kwon, D. T. Adroja, D. J. Voneshen, K. Kim, T.-H. Jang, J.-H.
Park, K.-Y. Choi, and S. Ji, Nat. Phys. {\bf 13}, 1079 (2017).
\bibitem{Attila_8}
A. Banerjee, P. Lampen-Kelley, J. Knolle, C. Balz, A. A. Aczel,
B. Winn, Y. Liu, D. Pajerowski, J. Yan, C. A. Bridges, A. T.
Savici, B. C. Chakoumakos, M. D. Lumsden, D. A. Tennant,
R. Moessner, D. G. Mandrus, and S. E. Nagler, Quantum
Materials {\bf 3}, 8 (2018).
\bibitem{Attila_9}
Y. Kasahara, T. Ohnishi, Y. Mizukami, O. Tanaka, S. Ma, K.
Sugii, N. Kurita, H. Tanaka, J. Nasu, Y. Motome, T. Shibauchi,
and Y. Matsuda, Nature (London) {\bf 559}, 227 (2018).
\bibitem{Our_Work}
M. P. Kwasigroch, B. Dou\c{c}ot, and C. Castelnovo, Phys. Rev. B {\bf 95}, 134439, (2017).
\bibitem{Hermele}
M. Hermele, M.P.A. Fisher, L. Balents, Phys. Rev. B {\bf 69}, 064404 (2004).
\bibitem{Polyakov_1974}
A. M. Polyakov, Nuclear Physics B {\bf 120}, 429 (1977)
\bibitem{Attila_36}
G. Chen, Phys. Rev. B {\bf 94}, 205107 (2016).
\bibitem{Attila_37}
G. Chen, Phys. Rev. B {\bf 96}, 195127 (2017).
\bibitem{Shannon}
O. Benton, O. Sikora, and N. Shannon, Phys. Rev. B {\bf 86}, 075154
(2012).
\bibitem{Attila_31}
N. Shannon, O. Sikora, F. Pollmann, K. Penc, and P. Fulde,
Phys. Rev. Lett. {\bf 108}, 067204 (2012).
\bibitem{Attila_32}
Y. Kato and S. Onoda, Phys. Rev. Lett. {\bf 115}, 077202 (2015).
\bibitem{Attila_34}
A. Banerjee, S. V. Isakov, K. Damle, and Y. B. Kim, Phys. Rev.
Lett. {\bf 100}, 047208 (2008).
\bibitem{Attila_35}
J.-P. Lv, G. Chen, Y. Deng, and Z. Y. Meng, Phys. Rev. Lett.
{\bf 115}, 037202 (2015).
\bibitem{Huang}
C. Huang, Y. Deng, Y. Wan, and Z. Y. Meng,
Phys. Rev. Lett. {\bf120}, 167202 (2018).
\bibitem{Attila_40}
L. Savary and L. Balents, Phys. Rev. Lett. {\bf 108}, 037202 (2012).
\bibitem{Attila_41}
S. B. Lee, S. Onoda, and L. Balents, Phys. Rev. B {\bf 86}, 104412
(2012).
\bibitem{Attila_42}
 L. Savary and L. Balents, Phys. Rev. B {\bf 87}, 205130 (2013).
\bibitem{Attila_43}
Z. Hao, A. G. R. Day, and M. J. P. Gingras, Phys. Rev. B {\bf 90},
214430 (2014).
\bibitem{Attila_44}
O. Benton, L. D. C. Jaubert, R. R. P. Singh, J. Oitmaa, and N.
Shannon, Phys. Rev. Lett. {\bf 121}, 067201 (2018).
\bibitem{Attila}
A. Szab\'o and C. Castelnovo, Phys. Rev. B {\bf 100}, 014417 (2019).
\bibitem{Polyakov_1978}
 A. M. Polyakov, Phys. Lett. B {\bf 72}, 477 (1978).
\bibitem{Villain}
J. Villain, J. Phys. {\bf 35}, 27 (1974).
\bibitem{Faddeev-Popov}
L.D. Faddeev and V. N. Popov, Phys. Lett. B {\bf 25}, 29 (1967).
\bibitem{Kosterlitz_1977}
J. M. Kosterlitz, J. Phys. C {\bf 10}, 3753 (1977).
\bibitem{spin-flip}
T. Fennell, P. P. Deen, A. R. Wildes, K. Schmalzl, D. Prabhakaran, A. T. Boothroyd, R. J. Aldus, D. F. McMorrow and S. T. Bramwell, Science {\bf 326}, 415 (2009).
\bibitem{Attila_73}
D. Khomskii, Nat. Commun. {\bf 3}, 904 (2012).
\bibitem{Attila_74}
E. Lantagne-Hurtubise, S. Bhattacharjee, and R. Moessner,
Phys. Rev. B {\bf 96}, 125145 (2017).
\bibitem{Attila_75}
 S. Nakosai and S. Onoda, J. Phys. Soc. Jpn. {\bf 88}, 053701 (2019).




























 
\end{thebibliography}
\end{document}